\documentclass[aps,pre,twocolumn,groupedaddress,superscriptaddress,showpacs]{revtex4-1}
\usepackage[utf8]{inputenc}
\usepackage{amsmath}
\usepackage{graphicx}
\usepackage{float}
\usepackage{xcolor}
\usepackage{cancel}
\usepackage{ulem}

\begin{document}

  \title {Rigid m-percolation in limited-valence gels}

\author{J. C. Neves}
\email{jlneves@ciencias.ulisboa.pt }
\affiliation{Centro de F\'{i}sica Te\'{o}rica e Computacional, Faculdade de Ci\^{e}ncias, Universidade de Lisboa, 1749-016
Lisboa, Portugal}
\affiliation{Departamento de F\'{\i}sica, Faculdade de Ci\^{e}ncias,
Universidade de Lisboa, 1749-016 Lisboa, Portugal}

  \author{J. M. Tavares}
   \email{jmtavares@fc.ul.pt}
    \affiliation{Centro de F\'{i}sica Te\'{o}rica e Computacional, Faculdade de Ci\^{e}ncias, Universidade de Lisboa, 1749-016 Lisboa, Portugal}
    \affiliation{Instituto Superior de Engenharia de Lisboa, ISEL, Avenida Conselheiro Em\'{i}dio Navarro, 1  1950-062 Lisboa - Portugal}
    
\author{N. A. M. Araújo}
    \email{nmaraujo@fc.ul.pt}
\affiliation{Centro de F\'{i}sica Te\'{o}rica e Computacional, Faculdade de Ci\^{e}ncias, Universidade de Lisboa, 1749-016
Lisboa, Portugal}
\affiliation{Departamento de F\'{\i}sica, Faculdade de Ci\^{e}ncias,
Universidade de Lisboa, 1749-016 Lisboa, Portugal}

 \author{C. S. Dias}
  \email{csdias@fc.ul.pt}
\affiliation{Centro de F\'{i}sica Te\'{o}rica e Computacional, Faculdade de Ci\^{e}ncias, Universidade de Lisboa, 1749-016
Lisboa, Portugal}
\affiliation{Departamento de F\'{\i}sica, Faculdade de Ci\^{e}ncias,
Universidade de Lisboa, 1749-016 Lisboa, Portugal}

\begin{abstract}

Determining the onset of rigidity in gels is a fundamental challenge with significant practical implications across different applications. Limited-valence, or patchy-particle systems have %\sout{emerged as} 
proven to be a valuable model to study the relationship between microscopic interactions and macroscopic mechanical properties.
It has been suggested that the emergence of rigidity coincides with the formation of an infinitely spanning cluster of particles with at least three bonds.
%\sout{It has been suggested that the rigidity percolation in these systems %coincides with the $m=3$-percolation transition, where particles belong to %the same rigid cluster if they are connected by at least $m=3$ bonds.}
This work explores this hypothesis, its implications, and its broader applicability across a range of system parameters, by associating the emergence of rigidity with  $m$-percolation transition for $m=3$. 
%\sout{To study the properties of the $m-$percolation transition, we %develop a mean-field theoretical approach validated with numerical %simulations} 
The properties of $m-$percolation  are developed using a mean-field theoretical approach validated with numerical simulations,
%{\sout{Using this approach, We build}} 
and used to build phase and rigidity diagrams for different particle valences of both single-component systems and binary mixtures of patchy particles. 
%\sout{We observe that the difference} 
The difference between connectivity and rigidity percolation thresholds 
%\sout{reduces}
is found to reduce with increasing valence, providing an explanation for the challenges encountered in experimental attempts to distinguish isotropic connectivity percolation from the onset of rigidity. For binary mixtures, we found a robust minimum average valence, below which the gel is never rigid.

  \end{abstract}
\maketitle

\section{Introduction}
Identifying the conditions for the onset of rigidity in gels is challenging \cite{Zhang2019,Colombo2014a,Fenton2023,Lu2008,Kohl2016,Valadez-Perez2013,Richards2017,Blumenfeld2005,Broedersz2011,Gomez2024,Bantawa2023,Damavandi2022,Damavandi2022a}. In this context, several hypotheses have been proposed, based on diverse concepts such as random percolation, directed percolation, and isostatic percolation \cite{Berthier2019,Ellenbroek2011,Ellenbroek2015,Jacobs1995,SampaioFilho2018,Nabizadeh2024}. Experimental results suggest that rigidity is solely reached after the percolation of local isostatic configurations \cite{Tsurusawa2019}. Recent numerical simulations of limited-valence particles show that the rigidity threshold coincides with that of percolation of particles with at least three bonds \cite{Dias2023a}. Here, we describe this transition using a  generalized concept of percolation, known as $m$-percolation \cite{Reich1978} ($m=3$ for the rigidity case), and study its dependence, for models of limited-valence particles,  on particle valence, temperature, density, and composition.

Limited-valence particles have garnered significant attention due to their ability to form low-density gels \cite{Markova2014,Romano2010,Howard2019,Zaccarelli2006,Sciortino2017,Capelot2012,Zaccarelli2005,Kim2023,Wang2012,Russo2009,Smallenburg2013,Li2020}. 
The impact of valence and mixture composition on gel formation \cite{Dias2013b,DelasHeras2010,Syk2016,Dias2013,Swinkels2024,Liu2020a}, the influence of gravity on gel structure \cite{DelasHeras2013,Teixeira2021,Geigenfeind2017}, the growth dynamics of gels on substrates or interfaces \cite{Dias2013,Araujo2015,Dias2016,Araujo2017,Dias2015,Dias2018a,Fenton2018,Dias2014,Dias2014a}, the use of linkers as an extra control over gelation \cite{Tavares2020,Lei2020,Antunes2019,Lowensohn2019,VanDerMeulen2013,Xia2020,Angioletti-Uberti2014,Xiong2009,Dias2020a,Kang2023} have  all been extensively studied assuming that  gelation was associated with %\sout{ordinary} 
connectivity (m=1) percolation. 
These gels are characterized by the formation of sparse networks of interconnected particles, that exhibit exotic equilibrium and dynamical properties \cite{Ruzicka2011,Smallenburg2013a,Sacanna2013,Shah2013,VanBlaaderen2013,Dias2017,Ortiz2014,King2024,Nag2024,Dias2018b,Wolters2015,Russo2022,Feng2013,Gong2017,Bantawa2021,Dias2018,Cui2023,Lee2011,Chang2021,Russo2024}.
However, as became recently evident through the direct probe of their mechanical properties \cite{Tsurusawa2019}, connectivity percolation is a necessary but not sufficient condition to obtain rigidity. Therefore, it is now clear that connectivity  percolation, while describing  a dramatic change in connectivity, does not  explain the onset of rigidity and that new approaches are needed to understand this emergence of new mechanical properties \cite{Tsurusawa2020}.

%\sout{Our primary aim} 
 The primary aim of this work is to elucidate how the $m$-percolation \cite{Reich1978} for $m=3$, manifests across various properties of limited-valence gels. 
 %{\sout{We employ a} 
 A mean-field theoretical approach, validated by numerical simulations, %\sout{which allows us} 
 is employed to explore a wide range of limited-valence structures. This approach 
 %\sout{enables us to construct} 
  results in the calculation of phase diagrams and percolation diagrams,  both for single component systems with various particle valences and  for binary mixtures with various mean valences, covering a broad range of model parameters, which is impractical using only simulations. 
  %\sout{By doing so, we aim to establish connections with the existing %literature on equilibrium gels and provide insights into the location of %the mechanical stability region within these systems.} 
  These results establish connections with the existing literature on equilibrium gels and provide insights into the location of their mechanical stability regions. 

In Section~\ref{sec.model}, 
%\sout{we present} 
the model studied in the simulations (Section~\ref{sec.sims}) and the mean field approach to $m$-percolation (Section~\ref{sec.mf}), both for single component systems and binary mixtures, are presented. Section~\ref{sec.results} describes the main results. In Section~\ref{sec.mpercolation}, 
%\sout{we compare}
the results of the simulations for the onset of rigidity  
are compared with the results of the mean-field approach for the $m=3$ percolation transition, showing that they coincide. Section~\ref{sec.diagram} focuses on the dependence of the onset of rigidity on the thermodynamical properties, while Section~\ref{sec.mix} extends the results to binary mixtures. 
%\sout{We draw the conclusions}
The conclusions are drawn in Section~\ref{sec.conclusions}.

\section{Model}\label{sec.model}

\subsection{Simulations}\label{sec.sims}

\begin{figure}[t]
   \begin{center}
\includegraphics[width=1.0\columnwidth]{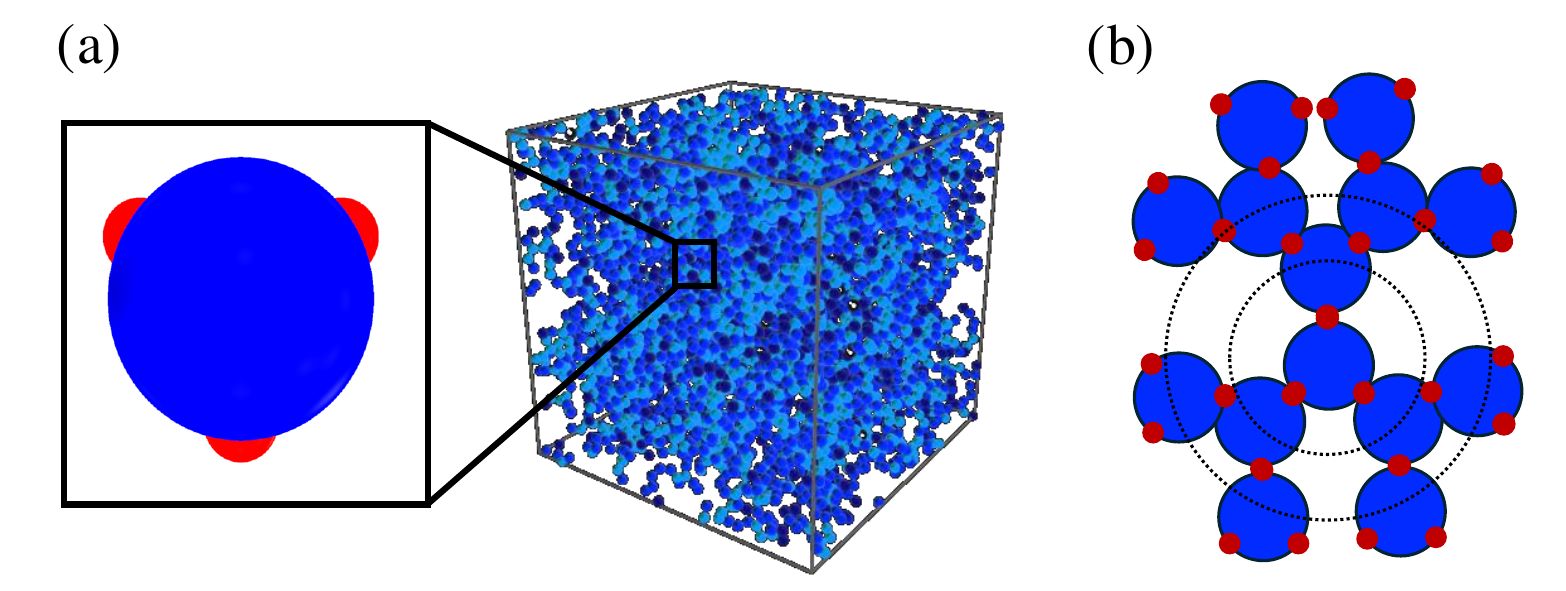} \\
\caption{\textbf{Model schematic representation.} Particles with valence three are used as an example. (a) Gel structure grown from a random distribution of three-valence colloids (right) and representation of the three-valence colloid model (left) with three attractive patches equidistant in the surface of the colloid. (b) Schematic representation of the mean-field assumption: clusters are tree like  and can be divided in levels (the thin circular lines identify the particles in the same level). } \label{fig.scheme}
   \end{center}
\end{figure}

We simulated a monodisperse suspension of spherical colloidal particles with radius $R$. To control the valence of each particle, we added point-like attractive sites equally spaced on their surfaces (see Fig.\ref{fig.scheme}(a)). Bonds between particles are only formed through these attractive sites, and the number of sites determines the valence of a particle \cite{Dias2016}. The pairwise interaction between particles was modeled as a superposition of an isotropic repulsion and a site-site attraction. 

The interaction between the cores is repulsive and is modeled by a pairwise potential,
\begin{equation}
U_{\text{colloid/colloid}}(r)=A/k e^{-k\left(r-2R\right)}, \label{eq.yukawa}
\end{equation}
where $R$ is the particle radius, $r$ is the distance between the centers, $A/k=0.25 k_BT$ sets the interaction strength, and  $k=0.8R^{-1}$ is the inverse screening length. The bonding interaction was modeled by an inverted Gaussian potential,
\begin{equation}
U_{\text{site/site}}(r_s)=-\epsilon e^{-(r_s/\sigma)^2}, \label{eq.gaussian} 
\end{equation}
where $r_s$ is the distance between the attractive sites, $\epsilon =20 k_BT$ represents the interaction strength, and $\sigma=0.2R$ is the width of the Gaussian. We obtained the trajectories numerically using Langevin dynamics and the LAMMPS open source library \cite{Plimpton1995}. The diffusion coefficients for translation, $D_\mathrm{t}$, and rotation, $D_\mathrm{r}$, are related by the Debye-Einstein relation
$D_\mathrm{r}=\frac{3}{4R^2}D_\mathrm{t}$~\cite{Mazza2007}.

To prepare the system for rheological measurements, we first minimized the energy through zero-temperature overdamped simulations. These simulations were run until the kinetic energy drops to at least ten orders of magnitude below its initial value. We then applied Lees-Edwards boundary conditions and imposed an oscillatory affine shear deformation, parametrized by 
\begin{gather}
 \mathbf{r'_\mathit{i}}=
  \begin{bmatrix}
   1 & \gamma(t) & 0 \\
   0 & 1 & 0 \\
   0 & 0 & 1 
   \end{bmatrix}
   \mathbf{r}_i, 
\end{gather}
where $\textbf{r}_i$ and $\textbf{r}'_i$ are the initial and final positions
of particle $i$, respectively, and $\gamma(t)=\gamma_0\sin\left(\omega t\right)$ is the shear strain. The linear viscoelastic regime is identified from the load curve, and we set $\gamma=0.07$ to focus on this regime. The first-order storage $G^{'}_1$  and loss $G^{''}_1$ moduli were then obtained from the time evolution of the shear stress $\sigma_{xy}(t)$, following the procedure outlined in Ref.~\cite{Colombo2014}.

\subsection{Mean Field theory for {\it{m}}-percolation}\label{sec.mf}

The concept of {\it{m}}-percolation (also called high density percolation) was introduced in \cite{Reich1978} and originates in the generalization of the definition of a cluster: two particles belong to the same {\it{m}}-cluster if they are bonded to each other and if each of them is bonded to at least $m$ particles. For $m=1$ the usual definition of a cluster is recovered; for $m>1$, only particles that belong to highly connected structures are considered. In this section the results for the {\it{m}}-percolation threshold and for the probability that a particle belongs to an infinite {\it{m}}-cluster \cite{Reich1978}
% this is defined in Reich1978
%\footnote{The probability that a particle belongs to an infinite $m$-%cluster is $P_m(\infty)\equiv \lim_{N\to \infty} N_m(N)/N$, where $N$ is %the number of particles and $N_m(N)$ is the number of particles that %belong to $m$-clusters with $N$ particles}
are derived for a single component system. The result for the m-percolation threshold is generalized to binary mixtures of particle species differing in their valence. The cases $m=1$ (connectivity percolation) and $m=3$ (rigidity percolation) will be dealt in more detail. The results are derived in a mean-field approach, where clusters are assumed to be tree like
%by treating the clusters in a mean field fashion: they are assumed to be %tree like 
(see Fig.~\ref{fig.scheme}(b)) and the bonds between the particles are assumed to be uncorrelated (random networks). The connectivity of the system is characterized by the bonding probability $p$ that, in the patchy particle models, is defined as the fraction of patches that are bonded, {\it{i.e.}}  the  number of bonds divided by the maximum possible number of bonds.

\subsubsection{{\it{m}}-percolation threshold}
The {\it{m}}-percolation threshold is the bonding probability $p_c(m)$ above which the probability of finding an infinite {\it{m}}-cluster becomes non-zero. 
The relation between $p_c(m)$ and the valence $f$ has been derived in \cite{Reich1978}. Here, we present a simpler derivation that can be generalized to binary mixtures.
The relation between $p_c(m)$ and the valence $f$ for the single component system of patchy particles under study, can be obtained by assuming that the clusters are tree like and that the bonds between the particles are independent and form with bonding probability $p$. Let us consider a tree like $m$-cluster,  divided in ordered  levels (see  Fig. \ref{fig.scheme}b), with  $n_i$ particles in level $i$. These particles all have at least $m$ bonds, from which at least $m-1$ will bond them to the particles of the next level, $i+1$. The number of particles $n_{i+1}$ in level $i+1$ of the cluster is then,
\begin{equation}
\label{ni+1ni}
n_{i+1}=n_i\sum_{k=m-1}^{f-1} k \binom{f-1}{k} p^k (1-p)^{f-1-k}.
\end{equation}
The percolation threshold $p_c(m)$ is the bonding probability for which the ratio $n_{i+1}/n_i$ is equal to 1 (see Eq.~(19) of \cite{Reich1978}):
\begin{equation}
\label{percm}
\sum_{k=m-1}^{f-1}k {{f-1}\choose{k}}p_c^k(1-p_c)^{f-1-k}=1.
\end{equation}
The usual percolation threshold is recovered for $m=1$, 
\begin{equation}
\label{pcm1}
p_c(1)=\frac{1}{f-1},
\end{equation}
and an analytical expression can also be obtained for the case $m=f$: $p_c(f)=[1/(f-1)]^{1/(f-1)}$. The expression for the $m=3$ percolation threshold is of particular importance, since, as stated before, it will be used as the threshold for the formation of a rigid gel,
\begin{equation}
\label{pcm3}
(f-1)p_c(3)\left[1-(1-p_c(3))^{f-2}\right]=1.
\end{equation}

The percolation thresholds $p_c(m)$ will be converted into percolation lines in the temperature ($T$) - density ($\rho$) phase diagram of the model under study, introduced in \ref{sec.sims},   using  Wertheim's theory as applied to this particular patchy particle model \cite{Dias2018b}.  The percolation lines (i.e. relations between density and temperatures) are obtained by solving together Eq.~\eqref{percm} and the law of mass action, 
\begin{equation}
\label{lma}
p_c=f \rho g_{HS}(\rho) \Delta(T) (1-p_c)^2,
\end{equation}
using the approximations for $g_{HS}(\rho)$ (the pair correlation function of hard spheres) and $\Delta(T)$ (the Mayer function of the pair potential defined by Eqs.~\eqref{eq.yukawa} and \eqref{eq.gaussian}) derived in  \cite{Dias2018b}. 

\subsubsection{Fraction of particles in infinite {\it{m}}-cluster's}

\begin{figure}[t]
\centering
\includegraphics[scale=0.45]{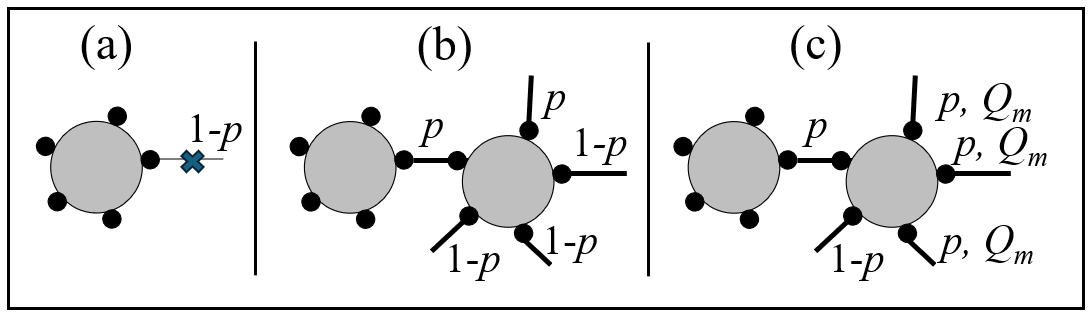}
\caption{{\textbf{Self consistent equation for $Q_m$.}} Schematic illustration  of the derivation of the self consistent Eq.~\ref{Qm} for  $Q_m$, the probability that a particle is not bonded to an infinite m-cluster through one of its patches. (a) the particle is not bonded through a given patch (see Eq.~\ref{Qa}); (b) the particle is bonded through the patch to another particle, but this particle has not  m or more patches bonded (see Eq.~\ref{Qb}); (c) the particle is bonded through the patch to another particle, this particle has m or more patches bonded, but none of these leads to an infinite m-cluster (see Eq.~\ref{Qc}).
}
\label{fig_calc}
\end{figure}

The fraction of particles in an infinite $m$-cluster was calculated in  \cite{Reich1978}, by using generating functions to obtain the   $m$-cluster size distribution function. Here we present an alternative and simpler derivation, that is a generalization  of the one used for $m=1$ in  polymer literature (e.g. \cite{Rubinstein2003}). 
This method relies on the definition of $Q_m$, the probability that a particle is not bonded to an infinite $m$-cluster through one of its patches, and on the derivation of a self consistent equation for $Q_m$. 
Let us consider one particle and one of its patches. There are 3 possibilities for this  patch that do not lead to an infinite m-cluster through it (see also figure \ref{fig_calc}):
\begin{itemize}
\item[(a)]{The patch is not bonded, which happens with probability,
\begin{equation}
\label{Qa}
Q_a=1-p;
\end{equation}
}
\item[(b)]{The patch is bonded to a particle which in turn is bonded to less than $m$ particles; this happens with probability,
\begin{equation}
\label{Qb}
Q_b=p \sum_{k=0}^{m-2}\binom{f-1}{k}p^k(1-p)^{f-1-k}.
\end{equation}
Notice that this probability is zero for $m<2$;
} 
\item[(c)]{The patch is bonded to a particle which in turn is bonded to more than $m$ particles, but none  of these bonds leads to an infinite m-cluster; this happens with probability,
\begin{equation}
\label{Qc}
  Q_c=p\sum_{k=m-1}^{f-1}\binom{f-1}{k}\left(Q_m-(1-p)\right)^k(1-p)^{f-1-k}.
\end{equation}
Notice that $(Q_m-(1-p))$ is the probability that a patch that is bonded does not lead to an infinite $m$-cluster. }
\end{itemize}
The probability $Q_m$ will then be implicitly defined as a function of $p$ and $f$ through,
\begin{equation}
\label{Qm}
Q_m=Q_a+Q_b+Q_c.
\end{equation}
Once $Q_m$ is determined, the probability that a particle belongs to an infinite $m$-cluster (or the fraction of particles that belong to an infinite $m$-cluster), $P_\infty(m)$, can be calculated using,
\begin{equation}
\label{Pinf}
P_\infty(m)=P_a-P_b,
\end{equation}
where,
\begin{equation}
\label{Pa}
P_a=\sum_{k=m}^f\binom{f}{k} p^k(1-p)^{f-k},
\end{equation}
is the fraction of particles that have $m$ or more patches bonded, and,
\begin{equation}
\label{Pb}
P_b=\sum_{k=m}^f\binom{f}{k} \left(Q_m-(1-p)\right)^k(1-p)^{f-k},
\end{equation}
is the fraction of particles that have $m$ or more patches bonded, but are not bonded to an infinite $m$-cluster through any of these patches.

Notice that, like in the case $m=1$ treated in \cite{Rubinstein2003}, $Q_m=1$ (i.e. no bonds lead to an infinite cluster) is always a solution of Eq.~\eqref{Qm} for any $p$, and in such case $P_\infty(m)=0$ (no particles belong to the infinite m-cluster).  A second solution for $Q_m$ will appear for $p>p_c(m)$, when the solutions of Eq.~\eqref{Qm} bifurcate (at $Q_m=1$). If Eq.~\eqref{Qm} is rewritten as $F(p,Q_m)=0$, then the value of $p$ at which the solution bifurcates can be obtained from $\left(\frac{\partial F}{\partial Q_m}\right)_{Q_m=1}=0$. It can easily be shown that the solution of this equation is $p=p_c(m)$ as given by Eq.~\eqref{percm}. 

Let us calculate $P_\infty(m)$ for the particular cases $m=1$ and $m=3, f=3$, those that will be compared to simulation results.
For the case of connectivity or $m=1$ percolation, the results of \cite{Rubinstein2003} are recovered:
\begin{equation}
\label{Qm1}
Q_1=1-p+pQ_1^{f-1},
\end{equation}
and,
\begin{equation}
\label{P1inf}
P_{\infty}(1)=1-Q_1^f.
\end{equation}
For $f=3$, the dependence of $P_{\infty}(1)$ in $p$ is obtained explicitly \cite{Rubinstein2003}: $Q_1=(1-p)/p$ and $P_{\infty}(1)=1-\left(\frac{1-p}{p}\right)^3$ (for $p \ge 1/2$).

The solution for $Q_3$ when $f=3$ can also be obtained as an explicit  function of $p$; for $p\ge p_c(3)= 1/\sqrt{2}$, it is,
\begin{equation}
\label{Q3p}
Q_3=\frac{(2p+1)(1-p)}{p}. 
\end{equation}
 The dependence of $P_\infty(3)$ on $p$ for $f=3$ and $p\ge p_c(3)=1/\sqrt{2}$ is obtained by replacing Eq.~\eqref{Q3p} in Eq.~\eqref{Pinf} and setting $f=3$, 
\begin{equation}
\label{P3inf}
P_\infty(3)=p^3- \left(\frac{1-p^2}{p}\right)^3.
\end{equation}

\subsubsection{$m$-percolation threshold for binary mixtures}{\label{sec-theorybin}}

Connectivity or $m=1$ percolation in single component systems of patchy particles,  as seen in the previous subsection, is determined by the valence of the particles and can only be obtained for valence higher or equal to three \cite{Bianchi2006}.
However, in  binary mixtures of patchy particles \cite{Teixeira2017a}
the composition of the mixture, the different valence of the species and the interactions between different types of patches, may lead to extra control over $m=1$ percolation. 
In particular, it has been shown that it is possible to reach $m=1$ percolation and gel formation with fewer bonds than in the case of single component systems, if a  binary mixtures with mean valence lower than 3 (i.e. fixing valence two or one in one of the species)  is considered \cite{Bianchi2006,DelasHeras2011}.
 It is then convenient to extend the concept of $m$-percolation to binary mixtures, in order to investigate the conditions under which a rigid gel may be obtained,  specially 
when the mean valence is brought to values lower than three. 

 We will restrict this extension to the case where all the patches of both species are of the same type (i.e. interact with each other with the same potential): the only distinction between the two species of the mixture is their number of patches \cite{DelasHeras2011}.  
We generalize the result for the $m$-percolation threshold to the case of the following mixture: species 1 has $f_1$ patches $A_1$ and $x_1$ is the fraction of particles of this species;
species 2 has $f_2$ patches of type $A_2$ and $x_2=1-x_1$ is the fraction of particles of this species. The composition of the mixture can also be characterized by its mean valence $\langle f \rangle$ (i.e. the mean number of patches per particle): 
\begin{equation}
\label{meanf}
\langle f \rangle  = x_1f_1+x_2f_2.
\end{equation}
 Like in \cite{DelasHeras2011}, the percolation thresholds will be expressed as a function of the probabilities $p_{A_i\to A_j}$, defined as the ratio between the number of bonds between patches of specie $i$ and patches of specie $j$ and the total number of patches of species $i$.  $p_{A_i\to A_j}$  is then, by definition, the probability that a patch of a particle of species  $i$ is bonded to a patch of a particle of species $j$.

The conditions for connectivity or $m=1$ percolation in this model are obtained from the construction of a matrix $\tilde T$ that encodes the structure of the clusters:  the tree clusters are divided in levels and the relation between the number of particles of specie $i$ in level $k+1$, $n_{i,k+1}$, and the number of particles of the previous level $k$, is \cite{DelasHeras2011},
\begin{equation}
\label{matrixbinperc}
\begin{bmatrix}
n_{1,k+1} \\
n_{2,k+1}
\end{bmatrix}
= \tilde T_1
\begin{bmatrix}
n_{1,k} \\
n_{2,k}
\end{bmatrix},
\end{equation}
with,
\begin{equation}
\label{matrix}
\tilde T_1=
\begin{bmatrix}
    (f_1-1)p_{A_1\to A_1}  & (f_2-1)p_{A_2\to A_1}   \\
     (f_1-1)p_{A_1\to A_2} &  (f_2-1)p_{A_2\to A_2}
\end{bmatrix}.
\end{equation}
Each element $t_{ij}$ of matrix $\tilde T_1$ represents the mean number of bonds between a particle  of type $j$ of level $k+1$ and a particle of type $i$ in level $k$. 
The percolation threshold is obtained, as a function of $f_i$ and of $p_{A_i \to A_j}$, by equating to 1 the largest absolute value of the eigenvalues of matrix $\tilde T$.

The generalization to high density or $m$-percolation is almost straightforward. The new matrix $\tilde T_m$ that accounts only for particles in level $k$ that have at least $m$ bonds, $m-1$ of which are bonds with particles of level $k+1$, is,
\begin{equation}
\label{matrixTm}
\tilde T_m=
\begin{bmatrix}
    \sum_{n=m-1}^{f_1-1}  T_{11,n}  &  \sum_{k=m-1}^{f_2-1}  T_{12,n} \\
  &  \\
\sum_{k=m-1}^{f_1-1}  T_{21,n}  &  \sum_{k=m-1}^{f_2-1} T_{22,n}
\end{bmatrix},
\end{equation}
where
\begin{equation}
\label{qijn}
T_{ij,n}=\binom{f_j-1}{n}(1-p_{A_j})^{f_j-1-n} \sum_{s=0}^n s \binom{n}{s} p_{A_j\to A_i}^s p_{A_j \to A_l}^{n-s},
\end{equation}
with $l\ne i$ and where $p_{A_j}\equiv p_{A_j\to A_j}+p_{A_j\to A_i}$ $(i\ne j )$ is the probability that a patch of species $j$ is bonded.   $T_{ij,n}$ is the mean number of bonds between a particle of type $j$ and particles of type $i$ in the next level of the cluster, when there are $n\ge m-1$ bonds between the particle $j$ and the  particles of the next level. The $n$ bonds out of particle of type $j$ may be formed with particles of type $i$ or with particles of type $l \ne i$, and thus the sum in Eq.~\eqref{qijn}.  The sum in Eq.~\eqref{qijn} can be performed to obtain,
\begin{equation}
\label{qijn_fin}
T_{ij,n}=n\binom{f_j-1}{n}p_{A_j\to A_i} (1-p_{A_j})^{f_j-1-n} p_{A_j}^{n-1}.
\end{equation}
Notice that when $f_i<m$, column $i$ of $\tilde T_m$ is 0.
It can be easily verified that when $m=1$ or $m=2$, Eq.~\eqref{qijn_fin} leads to the transfer matrix $\tilde T$ of Eq.~\eqref{matrix}. 
In general, the high density percolation threshold for any given value of $m$, will be obtained as a function of $p_{A_i\to A_j}$, $p_{A_i}$  and $f_i$ by determining, just like in the case $m=1$, the conditions under which the largest absolute value of the eigenvalues of $\tilde T_m$ equals 1.  

The probabilities $p_{A_i\to A_j}$ are not all independent since, from their definition,  $f_i x_i p_{A_i\to A_j}= f_j x_j p_{A_j\to A_i}$. Moreover,
the existence of just one type of patches imposes other relations between these and other 
 probabilities \cite{DelasHeras2011}: 
\begin{itemize}
\item[1.]{The bonding probability $p$ (the probability that a patch is bonded) is independent of the species to which a patch belongs and so:
\begin{equation}
\label{ppAipAj}
p=p_{A_1}=p_{A_2};
%=p_{A_1\to A_1}+p_{A_1\to A_2}=p_{A_2\to A_2}+p_{A_2\to A_1}.
\end{equation}
}
\item[2.]{The probability that a patch is bonded to a patch of a given species is independent of the species of the first patch:
\begin{equation}
\label{pAiAj}
p_{A_i\to A_j}=p_{A_j\to A_j} \qquad (i\ne j);
\end{equation}
}
\item[3.]{As a consequence of Eqs.~\eqref{ppAipAj} and \eqref{pAiAj}, the probabilities $p_{A_i\to A_j}$ can be expressed as a function of the probability $p$,  $x_i$ and $f_i$:
\begin{equation}
\label{pijsimp}
\begin{split}
&p_{A_1 \to A_1}=p_{A_2 \to A_1} = \frac{f_1 x}{\langle f\rangle } p, \\
&p_{A_2\to A_2}=p_{A_1 \to A_2} = \frac{f_2 x_2}{\langle f \rangle}p.
\end{split}
\end{equation}
}
%\item[4.]{}
\end{itemize}

It can be shown that the  matrix $\tilde T_m$ in Eq.~\eqref{matrixTm} has one zero eigenvalue and one positive eigenvalue.
The $m$-percolation thresholds  are determined by setting to one the non zero eigenvalue of $\tilde T_m$ \cite{DelasHeras2011}. The bonding probability $p_c(m)$ above which infinite $m$-clusters are formed is  then expressed as a function of $\langle f \rangle$ (or the compositions $x_i$), $f_1$ and $f_2$, using  Eqs.~\eqref{meanf}, \eqref{ppAipAj}, \eqref{pAiAj}, and \eqref{pijsimp}.
For $m=1$, the percolation threshold is
\begin{equation}
\label{pcbinm1}
p_c(1)=\frac{\langle f \rangle}{\langle f \rangle (f_1+f_2-1)-f_1f_2}.
\end{equation}
In the case $m=3$ the elements $t_{ij}(m=3)$ of the matrix in Eq.~\ref{matrixTm} are
\begin{equation}
\label{tijm3}  
t_{ij}(m=3)=\frac{f_i x_i p}{\langle f \rangle} (f_j-1)\left(1-(1-p)^{f_j-2}\right).
\end{equation} 
After setting to 1 the non-zero eigenvalue of $\tilde  T_3$, the percolation threshold for $m=3$, $p_c(m=3)$, is obtained as an implicit function of $\langle f \rangle$, $x_1$, $x_2$, $f_1$ and $f_2$,
\begin{equation}
\label{pcbinm3}
\frac{p_c(3)}{\langle f \rangle}\sum_{i=1,2}x_i f_i (f_i-1)\left[1-(1-p_c(3))^{f_i-2}\right]=1.
\end{equation}

The binary mixture in which one of the species has valence 2 is of special interest, since in this case it is possible to obtain  mean valences lower than 3 \cite{Bianchi2006,DelasHeras2011}. The expressions for the m-percolation thresholds $p_c(1)$ and $p_c(3)$  when $f_2=2$ and $f_1=f$ are,
\begin{equation}
\label{pcm1f22}
p_c(1)=\frac{\langle f \rangle}{\langle f \rangle (f+1)-2f},
\end{equation}
and,
\begin{equation}
\label{pcm3f22}
p_c(3)\frac{f(f-1)(\langle f \rangle -2)}{\langle f \rangle(f-2)}\left[1-(1-p_c(3))^{f-2}\right]=1.
\end{equation}

\begin{figure}[htb]
   \begin{center}
\includegraphics[width=1.0\columnwidth]{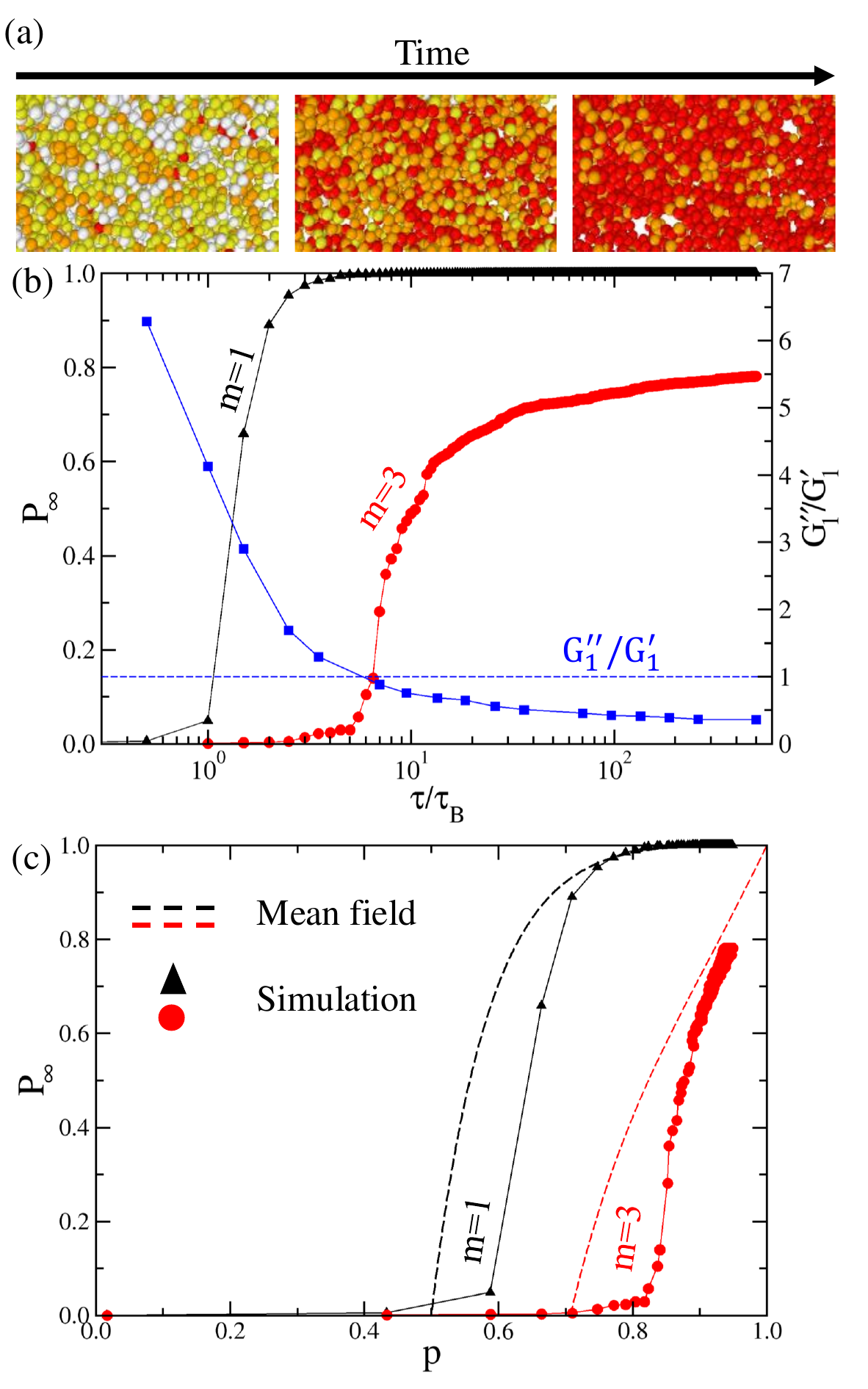} \\
\caption{\textbf{m-percolation rigidity transition.} (a) Snapshots of a valence $f=3$ particle system evolving over time and forming a gel. Colors represent the number of bonds: 0 (white), 1 (yellow), 2 (orange), and 3 (red). (b) Fraction of particles belonging to the largest cluster as a function of time for the $m=1$ (ordinary) percolation (black triangles), $m=3$ (three-bond) percolation (red circles), and the loss factor $\tan(\delta)=G^{\prime\prime}_1/G^{\prime}_1$  (blue triangles) as a function of time. (b) Fraction of particles belonging to the largest cluster as a function of the bonding probability $p$ for $m=1$ percolation, (black triangles) and $m=3$ percolation (red circles). Mean-field curves, obtained from Eq.~(\ref{P1inf}) for $m=1$ and from Eq.~(\ref{P3inf}) for $m=3$,
are shown in dashed lines. Simulations where performed for a 3D system of lateral size L=64R.} \label{fig.rigidity}
   \end{center}
\end{figure}

\begin{figure}[htb]
   \begin{center}
\includegraphics[width=1.0\columnwidth]{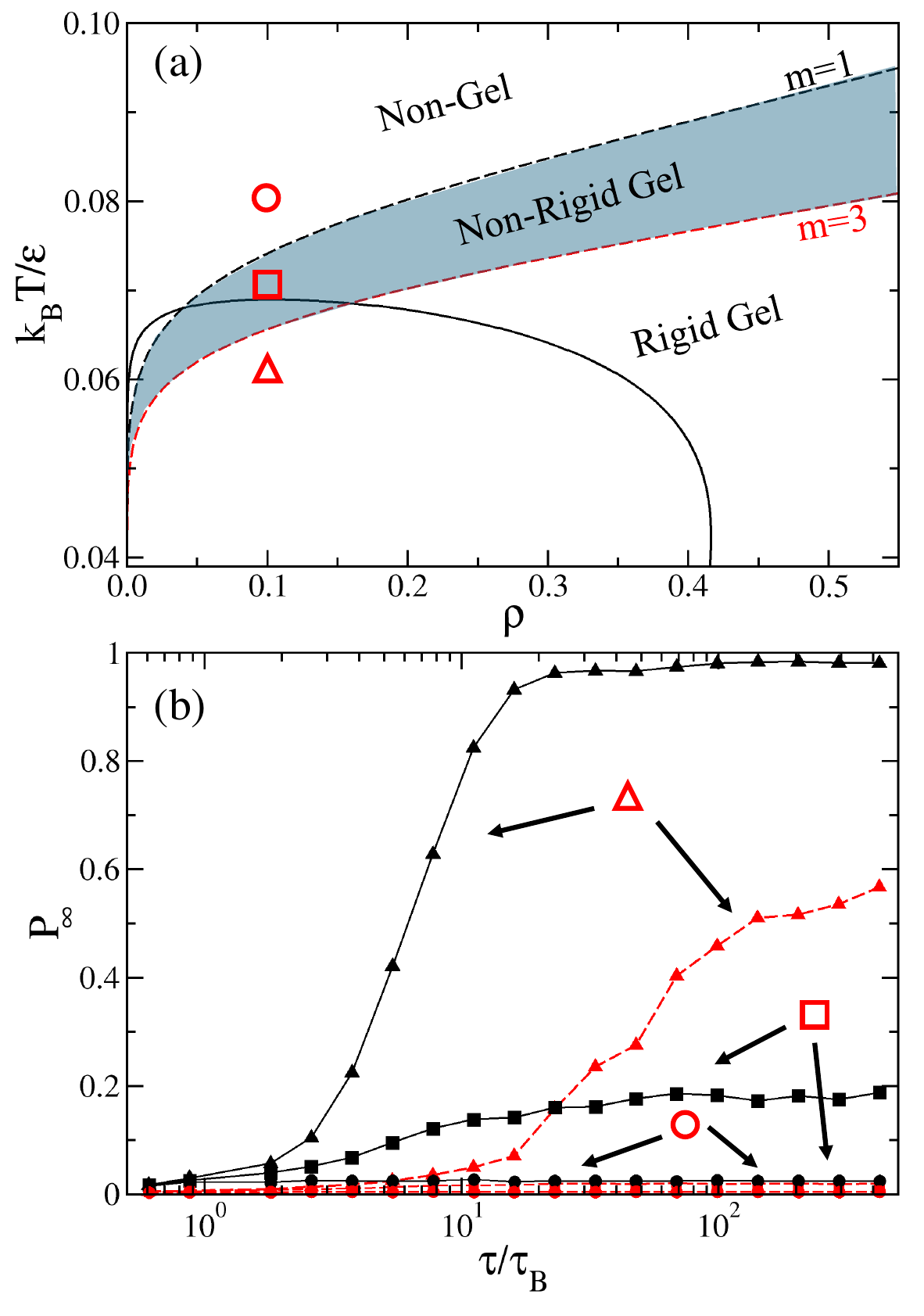} \\
\caption{\textbf{Phase and rigidity diagram.} (a) Temperature-density diagram for particles of valence $f=3$. The solid line indicates the liquid-gas coexistence curve. The top (black) dashed line is the $m=1$ percolation line and the bottom (red) dashed line is the $m=3$ percolation line. The shaded (blue) region between percolation transitions is the region where the formation of a non-rigid gel occurs. Above that region no gel is formed and below the gel is rigid. The three symbols indicate the numerical simulations for three different temperatures. (b) Fraction of particles belonging to the largest  m-cluster of particles for $m=1$ (black) and for $m=3$ (red) for a system of number density $\rho=0.1$. Simulations where performed for a 3D system of lateral size L=64R and averaged over 10 samples.} \label{fig.diagram}
   \end{center}
\end{figure}

\section{Results}\label{sec.results}

\subsection{{\it{m}}-percolation and rigidity}\label{sec.mpercolation}

We begin by growing a limited valence gel over time, starting with a random suspension of valence $f=3$ particles, as described in Ref.\cite{Dias2018b}. As shown in Fig.~\ref{fig.rigidity}(a), the number of bonds per particle evolves over time, leading to the formation of a fully connected structure.  To study $m$-percolation and its relation to the onset of rigidity, we plot in Fig.~\ref{fig.rigidity}(b) the fraction of particles belonging to the largest $m$-cluster, $P_{\infty}$, as a function of time for both $m=1$ and $m=3$. The mechanical rigidity of the gel is evident from the fact that $G^{\prime}_1$ is greater than $G^{\prime\prime}_1$, i.e. when the loss factor $G^{\prime\prime}_1$/$G^{\prime}_1<1$. This is illustrated in Fig.~\ref{fig.rigidity} by plotting the loss factor, which goes below one around the same point where the $m=3$ percolation transition occurs.

In Fig.~\ref{fig.rigidity}(c), we show the fraction of particles in the largest cluster, $P_{\infty}$, for both $m=1$ and $m=3$ clusters, as a function of the bonding probability, $p$. We also include the mean-field prediction of Eqs.~\eqref{P1inf} and \eqref{P3inf} for comparison with simulation results.  The mean-field prediction shows qualitative agreement with the simulations, supporting the validity of this approach for studying $m$-percolation in limited-valence systems.  However, as expected, there are quantitative deviations between the mean-field approach and the simulations. These deviations can be attributed to two factors: the presence of a small fraction of loops in the simulated gel, which is expected to increase the percolation threshold $p_c(m)$ in the simulations; the assumption, implicit in the comparison represented in Fig.~\ref{fig.rigidity}(c), that the simulation progresses through equilibrium intermediate states.

\begin{figure}[t]
   \begin{center}
\includegraphics[width=1.0\columnwidth]{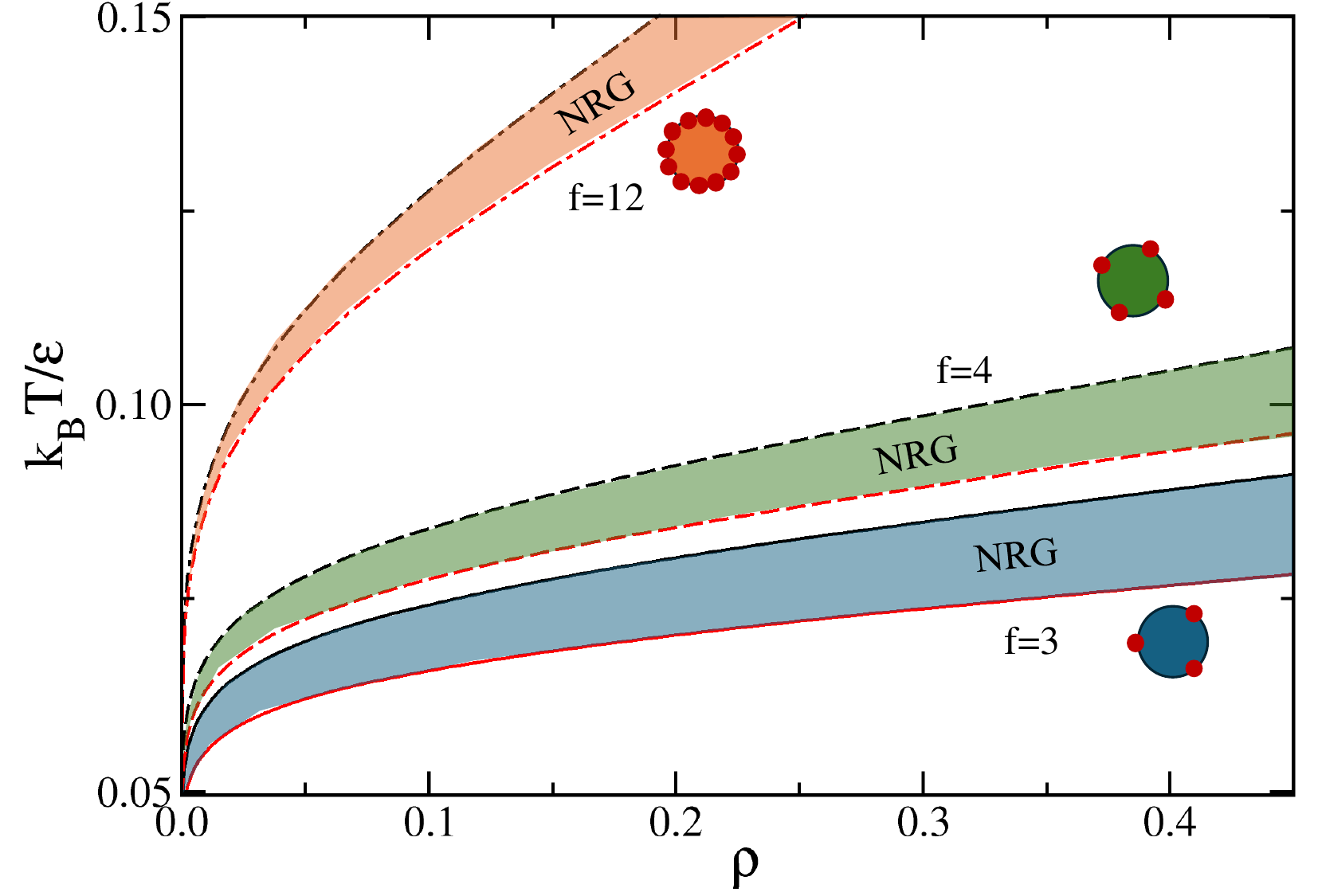} \\
\caption{\textbf{Valence dependence of the rigidity diagram.} Temperature-density diagram for particles of valence $f=\left\{3,4,12\right\}$. Shaded region indicate the Non-Rigid Gel (NRG) region between the $m=1$ (black dashed) percolation line and the $m=3$ (red dashed) percolation line. Top (orange) NRG region if for $f=12$, center (green) NRG region is for $f=4$, and bottom (blue) NRG region is for $f=3$.} \label{fig.valence}
   \end{center}
\end{figure}

\begin{figure}[t]
   \begin{center}
\includegraphics[width=0.9\columnwidth]{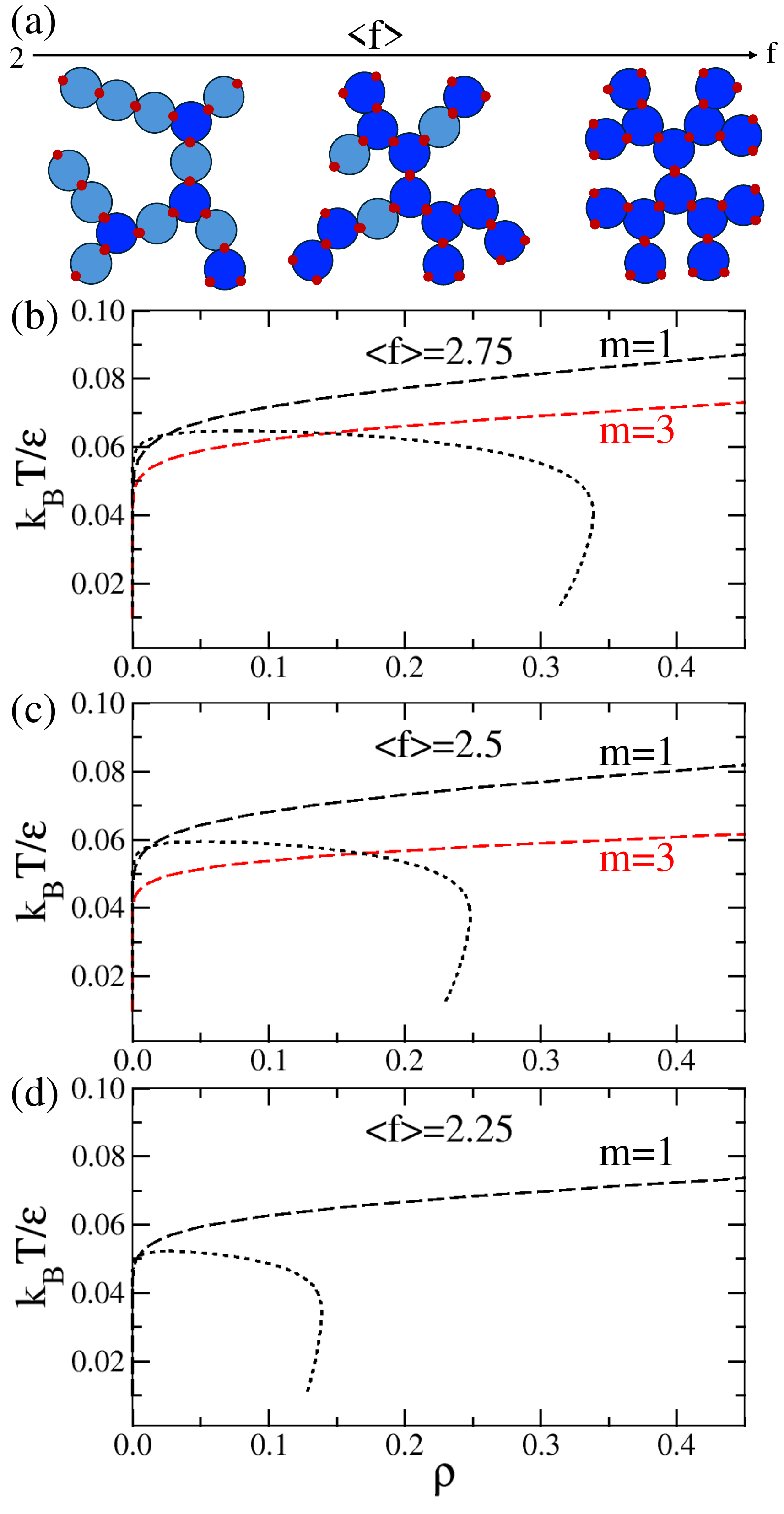} \\
\caption{\textbf{Phase and rigidity diagram of binary mixtures.} (a) Schematic representation of a binary mixture of particles with valence $f=3$ and  of particles with valence $f=2$. From left to right the mean valence $\langle f \rangle$ (see Eq.~\eqref{meanf}) increases. Phase and rigidity diagram for (b) $<f>=2.75$, (c) $<f>=2.5$, and (d) $<f>=2.25$. The dotted lines are an approximation for the liquid-gas coexistence lines, that illustrate the shrinking of the two phase region when $\langle f\rangle$ decreases (see the main text). The top (black) dashed line is the $m=1$ percolation line and the bottom (red) dashed line is the $m=3$ percolation line.} \label{fig.mixture}
   \end{center}
\end{figure}

\subsection{Phase Diagram}\label{sec.diagram}

In Fig.~\ref{fig.diagram}(a), we present the temperature versus density liquid-gas phase diagram for particles with valence $f=3$ as in Ref. \cite{Dias2018b}. The liquid-gas coexistence curve is depicted as a solid line and is calculated using Wertheim's first order perturbation theory with the approximations described in \cite{Dias2018b}.
The $m=1$ percolation transition, corresponding to particles with at least one bond, is shown as the top (black) dashed line, while the $m=3$ percolation transition, representing particles with at least three bonds, is shown as the bottom (red) dashed line. These lines are calculated using Eqs.~\eqref{lma} and \eqref{pcm1} for $m=1$ and Eqs.~\eqref{lma} and \eqref{pcm3} for $m=3$.
Notably, the theory predicts  the existence of a non rigid gel-like structure in a region of the phase diagram between these two lines.  The temperature range of this region increases with increasing particle density, making the distinction between the two percolation transitions more pronounced. 

This prediction of the theory is validated by the results of the numerical simulations. Gels of particles with valence $f=3$ were grown at number density $\rho=0.1$ and at 3 different temperatures $k_{\rm{B}}T/\epsilon=0.08$, 0.07 and 0.06; the time evolution of $P_\infty(m)$ ($m=1,3$) along these simulations is depicted in Fig.~\ref{fig.diagram}(b).  The theory predicts (see the symbols in Fig.~\ref{fig.diagram}(a)) the formation of a mechanically stable gel at the lowest temperature (triangle), the formation of a non-rigid gel at the intermediate temperature (square) and the absence of a gel at the highest temperature (circle).
The  simulation results for $P_{\infty}(\tau/\tau_{\rm {B}}\to \infty,m)$ in Fig.~\ref{fig.diagram}(b) confirm the existence (and location on the phase diagram) of the 3 gel regimes predicted by the theory:  at the lowest temperature  $P_{\infty}(\tau/\tau_{\rm {B}}\to \infty,3)$ is non zero, revealing that an infinite and rigid $m=3$ cluster was formed; at the intermediate temperature only $P_{\infty}(\tau/\tau_{\rm {B}}\to \infty,1)$ is non zero, an indication that a non rigid and infinite $m=1$ cluster was formed; finally, at the highest temperature, all $P_{\infty}(\tau/\tau_{\rm {B}}\to \infty,m) \approx 0$, and thus no gel has been formed.

In Fig.~\ref{fig.valence}, we present the rigidity diagram for particles with several valences. This temperature-density diagram features lines representing the $m=1$ percolation (dashed black) and the $m=3$ percolation (dashed red) for three different valences: $f=3$, $4$, and $12$. The region between the percolation lines, shaded and labeled as Non-Rigid Gel (NRG), highlights the range of conditions where a gel can form but lacks mechanical stability.  As the valence of the particles increases, the NRG region shifts to higher temperatures and becomes narrower. This observation offers a plausible explanation for the experimental challenge in distinguishing ordinary $m=1$ percolation from the onset of rigidity in gels. In the continuum limit, effectively represented by particles with a valence of 12, the difference between the $m=1$ and $m=3$ percolation transitions becomes very small.

\subsection{Binary Mixtures}\label{sec.mix}

The study of binary mixtures will be restricted to the model described in  section \ref{sec-theorybin} in the case $f_1=2$ and $f_2=f$.
This specific combination of valences allows us to investigate the emergence of $m$-percolation when the mean valence $\langle f \rangle$ is varied between $f$ and 2, a value for which both $m=1$ percolation and  liquid gas phase separation are known to be absent \cite{Bianchi2006,DelasHeras2011}. 

In Figs.~\ref{fig.mixture}(b), (c), and (d), we present the phase and $m$-percolation diagrams for various values of $\langle f\rangle$ in the range $]2;3[$, in mixtures  with $f=3$. These phase diagrams are calculated using an approximation that neglects the entropy of mixing, and thus obtains phase coexistence between states with different densities but equal composition (or equal $\langle f \rangle$). This approximation allows to calculate the phase diagrams using Wertheim's theory for single component systems (see section \ref{sec.diagram}) with $f$ replaced by $\langle f\rangle$ \cite{Bianchi2006}. 
However, this gross approximation captures an important 
feature of the complete phase diagrams of  mixtures of particles with two and three equal patches (i.e. that include the entropy of mixing \cite{DelasHeras2011}),    and that we intended to highlight:
the shrinking and eventual disappearance of the phase coexistence region when $\langle f \rangle \to 2^+$. This shrinking  can be seen in Figs.
~\ref{fig.mixture}(b), (c), and (d), and is interpreted as the possibility of obtaining low density network liquids that could form equilibrium gels \cite{Bianchi2006,DelasHeras2011}.

In Fig.~\ref{fig.mixture}(b), with $\langle f \rangle=2.75$, both the coexistence region and the non-rigid gel  region closely resemble the case $f=3$ of a single component system (see  Fig.~\ref{fig.diagram}(a)). As $\langle f \rangle$ decreases, the coexistence region shrinks and the percolation lines shift to lower temperatures. However, the non-rigid gel region expands significantly because the threshold temperatures of $m=3$ percolation   decrease more rapidly than those of $m=1$ percolation at a given density. For  $<f>=2.25$, the phase coexistence region has  shrank considerably, low density liquids have emerged, but the $m=3$ percolation transition is absent and only non rigid gels are obtained (see Fig.~\ref{fig.mixture}(d)).Therefore, these results suggest that the shrinking of the phase diagram that leads to the emergence of low density equilibrium gels  is accompanied by the complete loss of rigidity of those gels.

\begin{figure}[t]
   \begin{center}
\includegraphics[width=0.9\columnwidth]{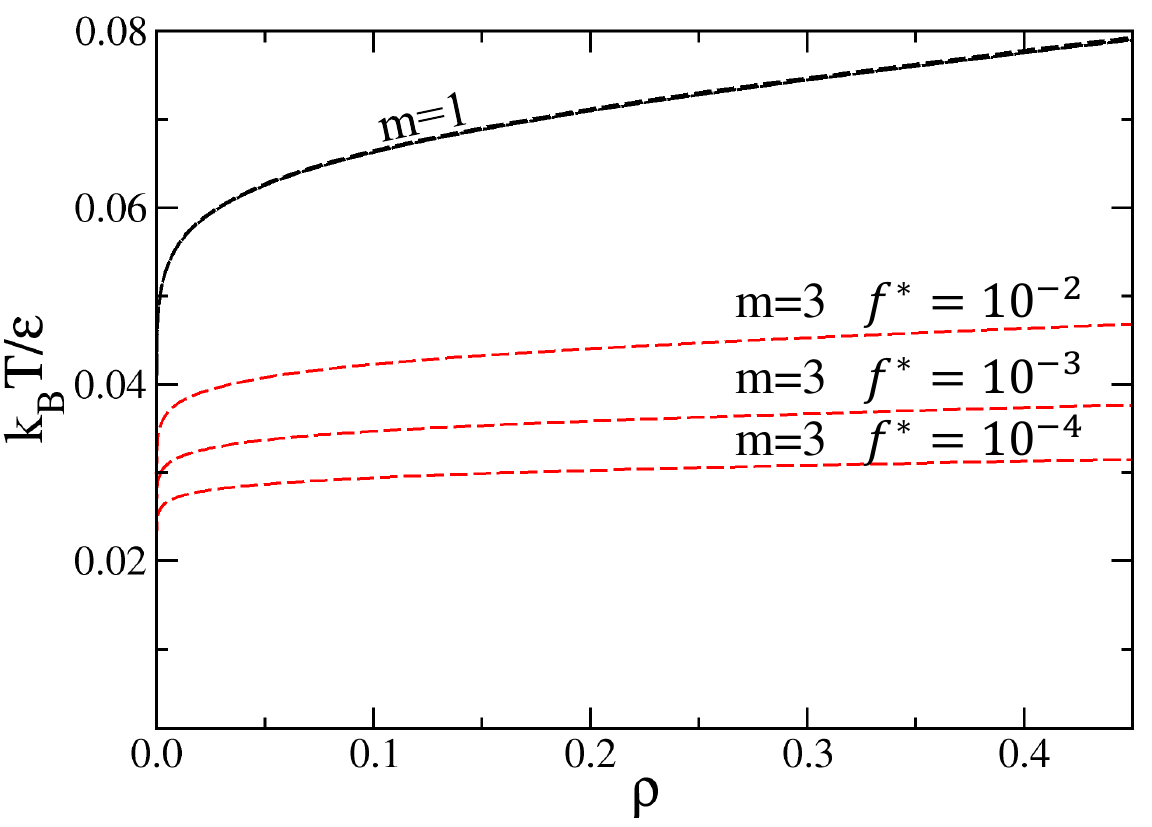} \\
\caption{\textbf{Rigidity diagram: approaching the vanish of  rigidity percolation.} Temperature-density rigidity diagram for binary mixtures of particles with 2 and $f=3$ patches. The top (black) dashed curve represents the $m=1$ percolation thresholds and the bottom (red) curves represent the $m=3$ percolation thresholds for the displayed values of $f^*=<f>-2.4$.   $\langle  f \rangle= 2.4$ is the limit below which, for these binary mixtures, the rigid gel ceases to exist.}
\label{fig.fstar}
   \end{center}
\end{figure}

The value of the mean valence $\langle f \rangle^*$ below which it becomes impossible to obtain a rigid gel in mixtures of particles with 2 and $f$ patches can be obtained by setting $p_c(3)=1$ in  (\ref{pcm3f22}),
\begin{equation}
\label{meanfth}
\langle f \rangle^*=\frac{2f(f-1)}{f(f-1)-f+2.}
\end{equation}
In the case $f=3$ this gives $\langle f \rangle^*=2.4$. To study the approach to this limit we have calculated m-percolation diagrams of a binary mixture of particles with 2 and $f=3$ patches for three low values of $f^*=\langle f \rangle -2.4$: $10^{-2}$, $10^{-3}$, and $10^{-4}$.
 As can be observed in Fig.\ref{fig.fstar}, the $m=1$ percolation line shows no significant change with $f^*$. On the other hand, $m=3$ percolation lines shift significantly to lower temperatures as $f^*$ is decreased, signaling a fast increase of the non rigid gel region, before the vanishing of the rigid gel at $\langle f \rangle=\langle f\rangle^*=2.4$. 

\begin{figure}[t]
   \begin{center}
\includegraphics[width=1.0\columnwidth]{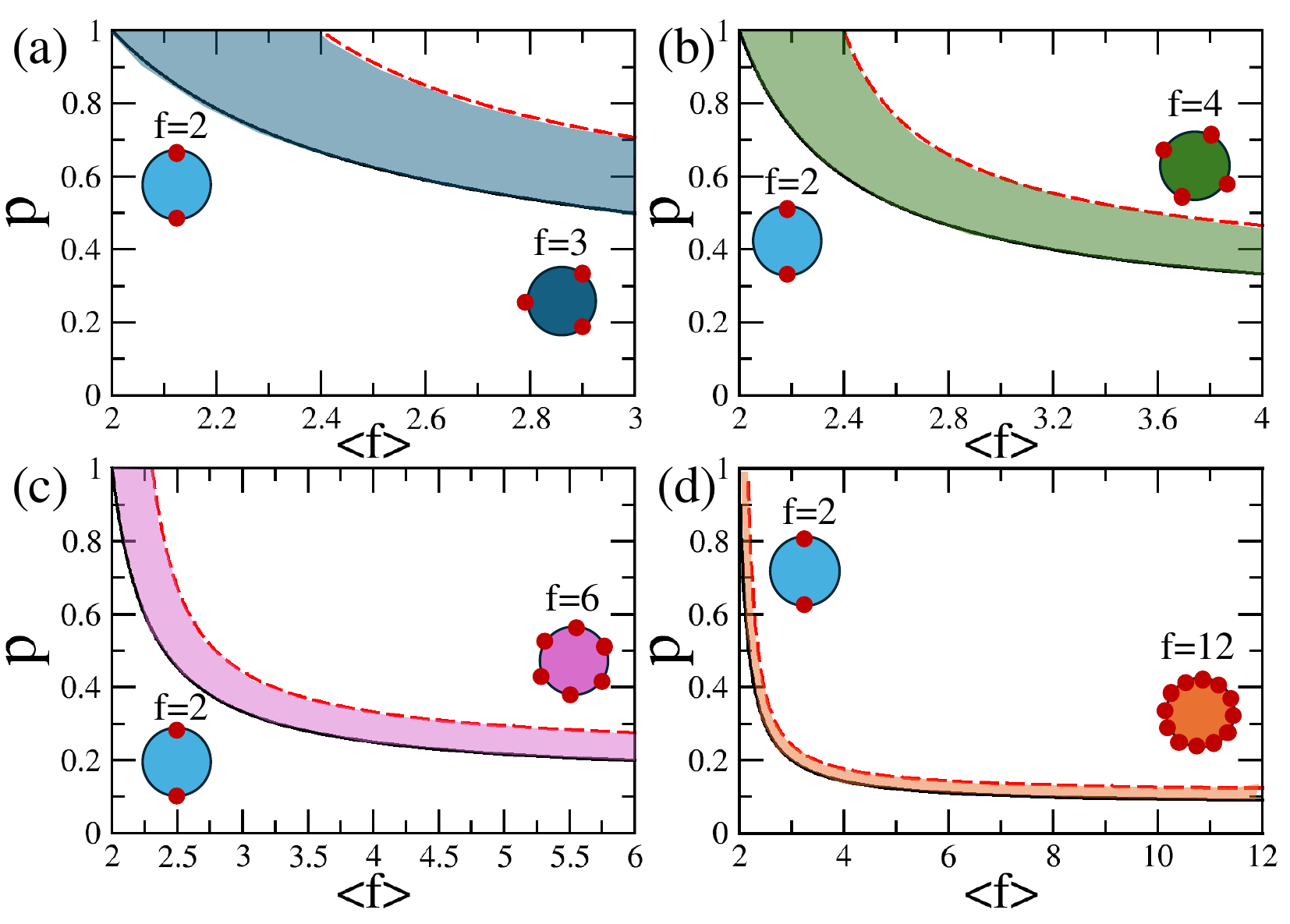} \\
\caption{\textbf{ Percolation diagrams of several binary mixtures.} Percolation thresholds $p_c(\langle f\rangle)$ for $m=1$ (black full lines) and $m=3$ (red dashed lines) of 4 distinct binary mixtures of particles with 2 and $f$ patches. The lines are calculated using Eqs.~\eqref{pcm1f22} and \eqref{pcm3f22}. (a) $f=3$, (b) $f=4$, (c) $f=6$, and (d) $f=12$. The rigid gel region is located above the red dashed line. The non-rigid gel region is located between the black full line and the red dashed line. Below the black full line the gel is absent.} \label{fig.mixvalence}
   \end{center}
\end{figure}

The $m=1$ and $m=3$ percolation thresholds of  four different binary mixtures of particles with 2 and $f$ patches ($f=3$, 4, 6 and 12), were calculated using Eqs.~\eqref{pcm1f22} and \eqref{pcm3f22}. The results are displayed
in Fig.\ref{fig.mixvalence}   as a bonding probability, $p$, vs mean functionality, $\langle f \rangle$, diagram. The range of $\langle f \rangle$ for which, at a given $p$, a non rigid gel exists decreases strongly with increasing $f$. This indicates (as was already established for the single component system - see Fig.~\ref{fig.valence}) that it becomes more difficult to distinguish non rigid and rigid gels when $f$ is increased.  It can also be observed that regions of these diagrams where a rigid gel exists are greatly increased when $f$ is increased. Finally, the limiting value $\langle f \rangle^*$ below which a rigid gel ceases to exist, while keeping  the value $\langle f \rangle^*=2.4$ for $f=3$ and 4, decreases fast to $2$ when $f$ increases.

\section{Conclusion}\label{sec.conclusions}

We investigated the onset of rigidity in gels using limited-valence patchy particle models. Our primary focus was to estimate the thresholds for the gel transitions in the phase diagram: from a liquid to a non-rigid gel and from a non-rigid gel to a rigid gel.
%We found evidence, from 
Simulations of a model with valence $f=3$ confirmed that the transition from a non-rigid to a rigid gel is due to the emergence of an infinite cluster where all particles form  3 bonds. The agreement between the simulation  and the mean-field theory for $m=3$ and $m=1$ percolation 
in  the location of these transitions on the phase diagram, led us to use and extend the theory to other limited valence models, including binary mixtures. 
The location of the gel transitions, liquid to non rigid, and non rigid to rigid, was  assumed to be associated with the emergence of $m=1$ and $m=3$ percolation, respectively.

The study of  single component systems of valence $f$ has revealed that the increase of $f$ leads to the broadening of the regions of the phase diagram where rigid gels can be found, and to the shrink of the regions of non-rigid gels. This same behavior was found in binary mixtures when the mean valence $\langle f \rangle$ was increased. However, the rigid gel can only exist for values of $\langle f \rangle$ above a certain threshold $\langle f \rangle^*$, whose value was predicted by the theory. 
The study of binary mixtures of particles with 2 and 3 equal patches has shown that the reported shrinking of the liquid gas phase separation in the limit $\langle f \rangle \to 2^+$ \cite{Bianchi2006,DelasHeras2011} is accompanied by a complete suppression of the rigid gel, that ceases to exist  when $\langle f \rangle  \le \langle f \rangle^*=2.4$.

The limited valence models studied in this work had  equal patches (i.e there was a single energy scale for bonding). However, it is known that both the phase diagrams and the $m=1$ percolation behavior of low valence patchy particle models are enriched when different types of patches are considered - with one \cite{Teixeira2021} or more \cite{Gouveia2022,Russo2011a,DelasHeras2012} energy scales for bonding. 
A generalization of the mean field m-percolation theory to these type of models could be used to investigate 
the possibility of overcoming the suppression of rigid gels at low mean valences.

It is important to notice that the mean field approach in this work assumes that they correspond to equilibrium configurations, whilst it is known that at least most gels (with the possible exception of the so called equilibrium gels \cite{Sciortino2017}, predicted by the theory for models with low valence at low densities) are inherently non-equilibrium systems. Future studies could explore non-equilibrium approaches or hybrid approaches combining non-equilibrium simulations with mean-field theory to further refine our understanding of gel rigidity.

Our findings have potential applications in various fields, particularly in bioengineering and materials science \cite{Lei2022}. The ability to control the density and mechanical response of gels is crucial for designing biocompatible scaffolds for tissue engineering and regenerative medicine \cite{Custodio2015,Custodio2014,Dias2020a}. The mechanical properties of these scaffolds can significantly influence cell behavior, including division and selectivity \cite{DeBelly2022,Sanz-Herrera2009,Vassaux2017,Gudipaty2017,Luo2013,Marinari2012}. Furthermore, this knowledge can be applied to the manufacturing of gels with tunable mechanical properties, a critical aspect for various industrial applications such as adhesives, coatings, and personal care products \cite{Mitura2020,Ioannidou2016,Banerjee2012,Coropceanu2022}.

\section*{Acknowledgments}

We acknowledge financial support from the Portuguese Foundation for Science and Technology (FCT) under Contracts no. PTDC/FIS-MAC/5689/2020 (https://doi.org/10.54499/PTDC/FIS-MAC/5689/2020), EXPL/FIS-MAC/0406/2021, CEECIND/00586/2017, UIDB/00618/2020 (DOI 10.54499/UIDB/00618/2020),  and UIDP/00618/2020 UIDP/00618/2020 (DOI 10.54499/UIDP/00618/2020).

\bibliography{library}% Produces the bibliography via BibTeX.

\begin{thebibliography}{105}%
\makeatletter
\providecommand \@ifxundefined [1]{%
 \@ifx{#1\undefined}
}%
\providecommand \@ifnum [1]{%
 \ifnum #1\expandafter \@firstoftwo
 \else \expandafter \@secondoftwo
 \fi
}%
\providecommand \@ifx [1]{%
 \ifx #1\expandafter \@firstoftwo
 \else \expandafter \@secondoftwo
 \fi
}%
\providecommand \natexlab [1]{#1}%
\providecommand \enquote  [1]{``#1''}%
\providecommand \bibnamefont  [1]{#1}%
\providecommand \bibfnamefont [1]{#1}%
\providecommand \citenamefont [1]{#1}%
\providecommand \href@noop [0]{\@secondoftwo}%
\providecommand \href [0]{\begingroup \@sanitize@url \@href}%
\providecommand \@href[1]{\@@startlink{#1}\@@href}%
\providecommand \@@href[1]{\endgroup#1\@@endlink}%
\providecommand \@sanitize@url [0]{\catcode `\\12\catcode `\$12\catcode `\&12\catcode `\#12\catcode `\^12\catcode `\_12\catcode `\%12\relax}%
\providecommand \@@startlink[1]{}%
\providecommand \@@endlink[0]{}%
\providecommand \url  [0]{\begingroup\@sanitize@url \@url }%
\providecommand \@url [1]{\endgroup\@href {#1}{\urlprefix }}%
\providecommand \urlprefix  [0]{URL }%
\providecommand \Eprint [0]{\href }%
\providecommand \doibase [0]{http://dx.doi.org/}%
\providecommand \selectlanguage [0]{\@gobble}%
\providecommand \bibinfo  [0]{\@secondoftwo}%
\providecommand \bibfield  [0]{\@secondoftwo}%
\providecommand \translation [1]{[#1]}%
\providecommand \BibitemOpen [0]{}%
\providecommand \bibitemStop [0]{}%
\providecommand \bibitemNoStop [0]{.\EOS\space}%
\providecommand \EOS [0]{\spacefactor3000\relax}%
\providecommand \BibitemShut  [1]{\csname bibitem#1\endcsname}%
\let\auto@bib@innerbib\@empty
%</preamble>
\bibitem [{\citenamefont {Zhang}\ \emph {et~al.}(2019)\citenamefont {Zhang}, \citenamefont {Zhang}, \citenamefont {Bouzid}, \citenamefont {Rocklin}, \citenamefont {Gado},\ and\ \citenamefont {Mao}}]{Zhang2019}%
  \BibitemOpen
  \bibfield  {author} {\bibinfo {author} {\bibfnamefont {S.}~\bibnamefont {Zhang}}, \bibinfo {author} {\bibfnamefont {L.}~\bibnamefont {Zhang}}, \bibinfo {author} {\bibfnamefont {M.}~\bibnamefont {Bouzid}}, \bibinfo {author} {\bibfnamefont {D.~Z.}\ \bibnamefont {Rocklin}}, \bibinfo {author} {\bibfnamefont {E.~D.}\ \bibnamefont {Gado}}, \ and\ \bibinfo {author} {\bibfnamefont {X.}~\bibnamefont {Mao}},\ }\href@noop {} {\bibfield  {journal} {\bibinfo  {journal} {Phys. Rev. Lett.}\ }\textbf {\bibinfo {volume} {123}},\ \bibinfo {pages} {058001} (\bibinfo {year} {2019})}\BibitemShut {NoStop}%
\bibitem [{\citenamefont {Colombo}\ and\ \citenamefont {Gado}(2014{\natexlab{a}})}]{Colombo2014a}%
  \BibitemOpen
  \bibfield  {author} {\bibinfo {author} {\bibfnamefont {J.}~\bibnamefont {Colombo}}\ and\ \bibinfo {author} {\bibfnamefont {E.~D.}\ \bibnamefont {Gado}},\ }\href@noop {} {\bibfield  {journal} {\bibinfo  {journal} {Soft Matt.}\ }\textbf {\bibinfo {volume} {10}},\ \bibinfo {pages} {4003} (\bibinfo {year} {2014}{\natexlab{a}})}\BibitemShut {NoStop}%
\bibitem [{\citenamefont {Fenton}\ \emph {et~al.}(2023)\citenamefont {Fenton}, \citenamefont {Padmanabhan}, \citenamefont {Ryu}, \citenamefont {Nguyen}, \citenamefont {Zia},\ and\ \citenamefont {Helgeson}}]{Fenton2023}%
  \BibitemOpen
  \bibfield  {author} {\bibinfo {author} {\bibfnamefont {S.~M.}\ \bibnamefont {Fenton}}, \bibinfo {author} {\bibfnamefont {P.}~\bibnamefont {Padmanabhan}}, \bibinfo {author} {\bibfnamefont {B.~K.}\ \bibnamefont {Ryu}}, \bibinfo {author} {\bibfnamefont {T.~T.~D.}\ \bibnamefont {Nguyen}}, \bibinfo {author} {\bibfnamefont {R.~N.}\ \bibnamefont {Zia}}, \ and\ \bibinfo {author} {\bibfnamefont {M.~E.}\ \bibnamefont {Helgeson}},\ }\href@noop {} {\bibfield  {journal} {\bibinfo  {journal} {Proc. Natl. Acad. Sci. U.S.A.}\ }\textbf {\bibinfo {volume} {120}},\ \bibinfo {pages} {e2215922120} (\bibinfo {year} {2023})}\BibitemShut {NoStop}%
\bibitem [{\citenamefont {Lu}\ \emph {et~al.}(2008)\citenamefont {Lu}, \citenamefont {Zaccarelli}, \citenamefont {Ciulla}, \citenamefont {Schofield}, \citenamefont {Sciortino},\ and\ \citenamefont {Weitz}}]{Lu2008}%
  \BibitemOpen
  \bibfield  {author} {\bibinfo {author} {\bibfnamefont {P.~J.}\ \bibnamefont {Lu}}, \bibinfo {author} {\bibfnamefont {E.}~\bibnamefont {Zaccarelli}}, \bibinfo {author} {\bibfnamefont {F.}~\bibnamefont {Ciulla}}, \bibinfo {author} {\bibfnamefont {A.~B.}\ \bibnamefont {Schofield}}, \bibinfo {author} {\bibfnamefont {F.}~\bibnamefont {Sciortino}}, \ and\ \bibinfo {author} {\bibfnamefont {D.~A.}\ \bibnamefont {Weitz}},\ }\href@noop {} {\bibfield  {journal} {\bibinfo  {journal} {Nature}\ }\textbf {\bibinfo {volume} {453}},\ \bibinfo {pages} {499} (\bibinfo {year} {2008})}\BibitemShut {NoStop}%
\bibitem [{\citenamefont {Kohl}\ \emph {et~al.}(2016)\citenamefont {Kohl}, \citenamefont {Capellmann}, \citenamefont {Laurati}, \citenamefont {Egelhaaf},\ and\ \citenamefont {Schmiedeberg}}]{Kohl2016}%
  \BibitemOpen
  \bibfield  {author} {\bibinfo {author} {\bibfnamefont {M.}~\bibnamefont {Kohl}}, \bibinfo {author} {\bibfnamefont {R.~F.}\ \bibnamefont {Capellmann}}, \bibinfo {author} {\bibfnamefont {M.}~\bibnamefont {Laurati}}, \bibinfo {author} {\bibfnamefont {S.~U.}\ \bibnamefont {Egelhaaf}}, \ and\ \bibinfo {author} {\bibfnamefont {M.}~\bibnamefont {Schmiedeberg}},\ }\href@noop {} {\bibfield  {journal} {\bibinfo  {journal} {Nat. Commun.}\ }\textbf {\bibinfo {volume} {7}},\ \bibinfo {pages} {11817} (\bibinfo {year} {2016})}\BibitemShut {NoStop}%
\bibitem [{\citenamefont {Valadez-Pérez}\ \emph {et~al.}(2013)\citenamefont {Valadez-Pérez}, \citenamefont {Liu}, \citenamefont {Eberle}, \citenamefont {Wagner},\ and\ \citenamefont {Castañeda-Priego}}]{Valadez-Perez2013}%
  \BibitemOpen
  \bibfield  {author} {\bibinfo {author} {\bibfnamefont {N.~E.}\ \bibnamefont {Valadez-Pérez}}, \bibinfo {author} {\bibfnamefont {Y.}~\bibnamefont {Liu}}, \bibinfo {author} {\bibfnamefont {A.~P.}\ \bibnamefont {Eberle}}, \bibinfo {author} {\bibfnamefont {N.~J.}\ \bibnamefont {Wagner}}, \ and\ \bibinfo {author} {\bibfnamefont {R.}~\bibnamefont {Castañeda-Priego}},\ }\href@noop {} {\bibfield  {journal} {\bibinfo  {journal} {Phys. Rev. E}\ }\textbf {\bibinfo {volume} {88}},\ \bibinfo {pages} {060302(R)} (\bibinfo {year} {2013})}\BibitemShut {NoStop}%
\bibitem [{\citenamefont {Richards}\ \emph {et~al.}(2017)\citenamefont {Richards}, \citenamefont {Hipp}, \citenamefont {Riley}, \citenamefont {Wagner},\ and\ \citenamefont {Butler}}]{Richards2017}%
  \BibitemOpen
  \bibfield  {author} {\bibinfo {author} {\bibfnamefont {J.~J.}\ \bibnamefont {Richards}}, \bibinfo {author} {\bibfnamefont {J.~B.}\ \bibnamefont {Hipp}}, \bibinfo {author} {\bibfnamefont {J.~K.}\ \bibnamefont {Riley}}, \bibinfo {author} {\bibfnamefont {N.~J.}\ \bibnamefont {Wagner}}, \ and\ \bibinfo {author} {\bibfnamefont {P.~D.}\ \bibnamefont {Butler}},\ }\href@noop {} {\bibfield  {journal} {\bibinfo  {journal} {Langmuir}\ }\textbf {\bibinfo {volume} {33}},\ \bibinfo {pages} {12260} (\bibinfo {year} {2017})}\BibitemShut {NoStop}%
\bibitem [{\citenamefont {Blumenfeld}\ \emph {et~al.}(2005)\citenamefont {Blumenfeld}, \citenamefont {Edwards},\ and\ \citenamefont {Ball}}]{Blumenfeld2005}%
  \BibitemOpen
  \bibfield  {author} {\bibinfo {author} {\bibfnamefont {R.}~\bibnamefont {Blumenfeld}}, \bibinfo {author} {\bibfnamefont {S.~F.}\ \bibnamefont {Edwards}}, \ and\ \bibinfo {author} {\bibfnamefont {R.~C.}\ \bibnamefont {Ball}},\ }\href@noop {} {\bibfield  {journal} {\bibinfo  {journal} {J. Phys. : Condens. Matter}\ }\textbf {\bibinfo {volume} {17}},\ \bibinfo {pages} {S2481} (\bibinfo {year} {2005})}\BibitemShut {NoStop}%
\bibitem [{\citenamefont {Broedersz}\ \emph {et~al.}(2011)\citenamefont {Broedersz}, \citenamefont {Mao}, \citenamefont {Lubensky},\ and\ \citenamefont {Mackintosh}}]{Broedersz2011}%
  \BibitemOpen
  \bibfield  {author} {\bibinfo {author} {\bibfnamefont {C.~P.}\ \bibnamefont {Broedersz}}, \bibinfo {author} {\bibfnamefont {X.}~\bibnamefont {Mao}}, \bibinfo {author} {\bibfnamefont {T.~C.}\ \bibnamefont {Lubensky}}, \ and\ \bibinfo {author} {\bibfnamefont {F.~C.}\ \bibnamefont {Mackintosh}},\ }\href@noop {} {\bibfield  {journal} {\bibinfo  {journal} {Nat. Phys.}\ }\textbf {\bibinfo {volume} {7}},\ \bibinfo {pages} {983} (\bibinfo {year} {2011})}\BibitemShut {NoStop}%
\bibitem [{\citenamefont {Gomez}\ and\ \citenamefont {Rovigatti}(2024)}]{Gomez2024}%
  \BibitemOpen
  \bibfield  {author} {\bibinfo {author} {\bibfnamefont {S.~S.}\ \bibnamefont {Gomez}}\ and\ \bibinfo {author} {\bibfnamefont {L.}~\bibnamefont {Rovigatti}},\ }\href@noop {} {\bibfield  {journal} {\bibinfo  {journal} {J. Chem. Phys.}\ }\textbf {\bibinfo {volume} {160}},\ \bibinfo {pages} {184901} (\bibinfo {year} {2024})}\BibitemShut {NoStop}%
\bibitem [{\citenamefont {Bantawa}\ \emph {et~al.}(2023)\citenamefont {Bantawa}, \citenamefont {Keshavarz}, \citenamefont {Geri}, \citenamefont {Bouzid}, \citenamefont {Divoux}, \citenamefont {McKinley},\ and\ \citenamefont {Gado}}]{Bantawa2023}%
  \BibitemOpen
  \bibfield  {author} {\bibinfo {author} {\bibfnamefont {M.}~\bibnamefont {Bantawa}}, \bibinfo {author} {\bibfnamefont {B.}~\bibnamefont {Keshavarz}}, \bibinfo {author} {\bibfnamefont {M.}~\bibnamefont {Geri}}, \bibinfo {author} {\bibfnamefont {M.}~\bibnamefont {Bouzid}}, \bibinfo {author} {\bibfnamefont {T.}~\bibnamefont {Divoux}}, \bibinfo {author} {\bibfnamefont {G.~H.}\ \bibnamefont {McKinley}}, \ and\ \bibinfo {author} {\bibfnamefont {E.~D.}\ \bibnamefont {Gado}},\ }\href@noop {} {\bibfield  {journal} {\bibinfo  {journal} {Nat. Phys.}\ }\textbf {\bibinfo {volume} {19}},\ \bibinfo {pages} {1178} (\bibinfo {year} {2023})}\BibitemShut {NoStop}%
\bibitem [{\citenamefont {Damavandi}\ \emph {et~al.}(2022{\natexlab{a}})\citenamefont {Damavandi}, \citenamefont {Hagh}, \citenamefont {Santangelo},\ and\ \citenamefont {Manning}}]{Damavandi2022}%
  \BibitemOpen
  \bibfield  {author} {\bibinfo {author} {\bibfnamefont {O.~K.}\ \bibnamefont {Damavandi}}, \bibinfo {author} {\bibfnamefont {V.~F.}\ \bibnamefont {Hagh}}, \bibinfo {author} {\bibfnamefont {C.~D.}\ \bibnamefont {Santangelo}}, \ and\ \bibinfo {author} {\bibfnamefont {M.~L.}\ \bibnamefont {Manning}},\ }\href@noop {} {\bibfield  {journal} {\bibinfo  {journal} {Phys. Rev. E}\ }\textbf {\bibinfo {volume} {105}},\ \bibinfo {pages} {025004} (\bibinfo {year} {2022}{\natexlab{a}})}\BibitemShut {NoStop}%
\bibitem [{\citenamefont {Damavandi}\ \emph {et~al.}(2022{\natexlab{b}})\citenamefont {Damavandi}, \citenamefont {Hagh}, \citenamefont {Santangelo},\ and\ \citenamefont {Manning}}]{Damavandi2022a}%
  \BibitemOpen
  \bibfield  {author} {\bibinfo {author} {\bibfnamefont {O.~K.}\ \bibnamefont {Damavandi}}, \bibinfo {author} {\bibfnamefont {V.~F.}\ \bibnamefont {Hagh}}, \bibinfo {author} {\bibfnamefont {C.~D.}\ \bibnamefont {Santangelo}}, \ and\ \bibinfo {author} {\bibfnamefont {M.~L.}\ \bibnamefont {Manning}},\ }\href@noop {} {\bibfield  {journal} {\bibinfo  {journal} {Phys. Rev. E}\ }\textbf {\bibinfo {volume} {105}},\ \bibinfo {pages} {025003} (\bibinfo {year} {2022}{\natexlab{b}})}\BibitemShut {NoStop}%
\bibitem [{\citenamefont {Berthier}\ \emph {et~al.}(2019)\citenamefont {Berthier}, \citenamefont {Kollmer}, \citenamefont {Henkes}, \citenamefont {Liu}, \citenamefont {Schwarz},\ and\ \citenamefont {Daniels}}]{Berthier2019}%
  \BibitemOpen
  \bibfield  {author} {\bibinfo {author} {\bibfnamefont {E.}~\bibnamefont {Berthier}}, \bibinfo {author} {\bibfnamefont {J.~E.}\ \bibnamefont {Kollmer}}, \bibinfo {author} {\bibfnamefont {S.~E.}\ \bibnamefont {Henkes}}, \bibinfo {author} {\bibfnamefont {K.}~\bibnamefont {Liu}}, \bibinfo {author} {\bibfnamefont {J.~M.}\ \bibnamefont {Schwarz}}, \ and\ \bibinfo {author} {\bibfnamefont {K.~E.}\ \bibnamefont {Daniels}},\ }\href@noop {} {\bibfield  {journal} {\bibinfo  {journal} {Phys. Rev. Mater.}\ }\textbf {\bibinfo {volume} {3}},\ \bibinfo {pages} {075602} (\bibinfo {year} {2019})}\BibitemShut {NoStop}%
\bibitem [{\citenamefont {Ellenbroek}\ and\ \citenamefont {Mao}(2011)}]{Ellenbroek2011}%
  \BibitemOpen
  \bibfield  {author} {\bibinfo {author} {\bibfnamefont {W.~G.}\ \bibnamefont {Ellenbroek}}\ and\ \bibinfo {author} {\bibfnamefont {X.}~\bibnamefont {Mao}},\ }\href@noop {} {\bibfield  {journal} {\bibinfo  {journal} {EPL}\ }\textbf {\bibinfo {volume} {96}},\ \bibinfo {pages} {54002} (\bibinfo {year} {2011})}\BibitemShut {NoStop}%
\bibitem [{\citenamefont {Ellenbroek}\ \emph {et~al.}(2015)\citenamefont {Ellenbroek}, \citenamefont {Hagh}, \citenamefont {Kumar}, \citenamefont {Thorpe},\ and\ \citenamefont {Hecke}}]{Ellenbroek2015}%
  \BibitemOpen
  \bibfield  {author} {\bibinfo {author} {\bibfnamefont {W.~G.}\ \bibnamefont {Ellenbroek}}, \bibinfo {author} {\bibfnamefont {V.~F.}\ \bibnamefont {Hagh}}, \bibinfo {author} {\bibfnamefont {A.}~\bibnamefont {Kumar}}, \bibinfo {author} {\bibfnamefont {M.~F.}\ \bibnamefont {Thorpe}}, \ and\ \bibinfo {author} {\bibfnamefont {M.~V.}\ \bibnamefont {Hecke}},\ }\href@noop {} {\bibfield  {journal} {\bibinfo  {journal} {Phys. Rev. Lett.}\ }\textbf {\bibinfo {volume} {114}},\ \bibinfo {pages} {135501} (\bibinfo {year} {2015})}\BibitemShut {NoStop}%
\bibitem [{\citenamefont {Jacobs}\ and\ \citenamefont {Thorpe}(1995)}]{Jacobs1995}%
  \BibitemOpen
  \bibfield  {author} {\bibinfo {author} {\bibfnamefont {D.~J.}\ \bibnamefont {Jacobs}}\ and\ \bibinfo {author} {\bibfnamefont {M.~F.}\ \bibnamefont {Thorpe}},\ }\href@noop {} {\bibfield  {journal} {\bibinfo  {journal} {Phys. Rev. Lett.}\ }\textbf {\bibinfo {volume} {75}},\ \bibinfo {pages} {4051} (\bibinfo {year} {1995})}\BibitemShut {NoStop}%
\bibitem [{\citenamefont {Filho}\ \emph {et~al.}(2018)\citenamefont {Filho}, \citenamefont {Jr.}, \citenamefont {Herrmann},\ and\ \citenamefont {Moreira}}]{SampaioFilho2018}%
  \BibitemOpen
  \bibfield  {author} {\bibinfo {author} {\bibfnamefont {C.~I. N.~S.}\ \bibnamefont {Filho}}, \bibinfo {author} {\bibfnamefont {J.~S.~A.}\ \bibnamefont {Jr.}}, \bibinfo {author} {\bibfnamefont {H.~J.}\ \bibnamefont {Herrmann}}, \ and\ \bibinfo {author} {\bibfnamefont {A.~A.}\ \bibnamefont {Moreira}},\ }\href@noop {} {\bibfield  {journal} {\bibinfo  {journal} {Phys. Rev. Lett.}\ }\textbf {\bibinfo {volume} {120}},\ \bibinfo {pages} {175701} (\bibinfo {year} {2018})}\BibitemShut {NoStop}%
\bibitem [{\citenamefont {Nabizadeh}\ \emph {et~al.}(2024)\citenamefont {Nabizadeh}, \citenamefont {Nasirian}, \citenamefont {Li}, \citenamefont {Saraswat}, \citenamefont {Waheibi}, \citenamefont {Hsiao}, \citenamefont {Bi}, \citenamefont {Ravandi},\ and\ \citenamefont {Jamali}}]{Nabizadeh2024}%
  \BibitemOpen
  \bibfield  {author} {\bibinfo {author} {\bibfnamefont {M.}~\bibnamefont {Nabizadeh}}, \bibinfo {author} {\bibfnamefont {F.}~\bibnamefont {Nasirian}}, \bibinfo {author} {\bibfnamefont {X.}~\bibnamefont {Li}}, \bibinfo {author} {\bibfnamefont {Y.}~\bibnamefont {Saraswat}}, \bibinfo {author} {\bibfnamefont {R.}~\bibnamefont {Waheibi}}, \bibinfo {author} {\bibfnamefont {L.~C.}\ \bibnamefont {Hsiao}}, \bibinfo {author} {\bibfnamefont {D.}~\bibnamefont {Bi}}, \bibinfo {author} {\bibfnamefont {B.}~\bibnamefont {Ravandi}}, \ and\ \bibinfo {author} {\bibfnamefont {S.}~\bibnamefont {Jamali}},\ }\href@noop {} {\bibfield  {journal} {\bibinfo  {journal} {Proc. Natl. Acad. Sci. U.S.A.}\ }\textbf {\bibinfo {volume} {121}},\ \bibinfo {pages} {e2316394121} (\bibinfo {year} {2024})}\BibitemShut {NoStop}%
\bibitem [{\citenamefont {Tsurusawa}\ \emph {et~al.}(2019)\citenamefont {Tsurusawa}, \citenamefont {Leocmach}, \citenamefont {Russo},\ and\ \citenamefont {Tanaka}}]{Tsurusawa2019}%
  \BibitemOpen
  \bibfield  {author} {\bibinfo {author} {\bibfnamefont {H.}~\bibnamefont {Tsurusawa}}, \bibinfo {author} {\bibfnamefont {M.}~\bibnamefont {Leocmach}}, \bibinfo {author} {\bibfnamefont {J.}~\bibnamefont {Russo}}, \ and\ \bibinfo {author} {\bibfnamefont {H.}~\bibnamefont {Tanaka}},\ }\href@noop {} {\bibfield  {journal} {\bibinfo  {journal} {Sci. Adv.}\ }\textbf {\bibinfo {volume} {5}},\ \bibinfo {pages} {eaav6090} (\bibinfo {year} {2019})}\BibitemShut {NoStop}%
\bibitem [{\citenamefont {Dias}\ \emph {et~al.}(2023)\citenamefont {Dias}, \citenamefont {Neves}, \citenamefont {da~Gama}, \citenamefont {del Gado},\ and\ \citenamefont {Araujo}}]{Dias2023a}%
  \BibitemOpen
  \bibfield  {author} {\bibinfo {author} {\bibfnamefont {C.~S.}\ \bibnamefont {Dias}}, \bibinfo {author} {\bibfnamefont {J.~C.}\ \bibnamefont {Neves}}, \bibinfo {author} {\bibfnamefont {M.~M.~T.}\ \bibnamefont {da~Gama}}, \bibinfo {author} {\bibfnamefont {E.}~\bibnamefont {del Gado}}, \ and\ \bibinfo {author} {\bibfnamefont {N.~A.~M.}\ \bibnamefont {Araujo}},\ }\href@noop {} {\bibfield  {journal} {\bibinfo  {journal} {arXiv:2307.13315}\ } (\bibinfo {year} {2023})}\BibitemShut {NoStop}%
\bibitem [{\citenamefont {Reich}\ and\ \citenamefont {Leath}(1978)}]{Reich1978}%
  \BibitemOpen
  \bibfield  {author} {\bibinfo {author} {\bibfnamefont {G.~R.}\ \bibnamefont {Reich}}\ and\ \bibinfo {author} {\bibfnamefont {P.~L.}\ \bibnamefont {Leath}},\ }\href@noop {} {\bibfield  {journal} {\bibinfo  {journal} {J. Stat. Phys.}\ }\textbf {\bibinfo {volume} {19}},\ \bibinfo {pages} {611} (\bibinfo {year} {1978})}\BibitemShut {NoStop}%
\bibitem [{\citenamefont {Markova}\ \emph {et~al.}(2014)\citenamefont {Markova}, \citenamefont {Alberts}, \citenamefont {Munro},\ and\ \citenamefont {Lenne}}]{Markova2014}%
  \BibitemOpen
  \bibfield  {author} {\bibinfo {author} {\bibfnamefont {O.}~\bibnamefont {Markova}}, \bibinfo {author} {\bibfnamefont {J.}~\bibnamefont {Alberts}}, \bibinfo {author} {\bibfnamefont {E.}~\bibnamefont {Munro}}, \ and\ \bibinfo {author} {\bibfnamefont {P.-F.}\ \bibnamefont {Lenne}},\ }\href@noop {} {\bibfield  {journal} {\bibinfo  {journal} {Phys. Rev. E}\ }\textbf {\bibinfo {volume} {90}},\ \bibinfo {pages} {022301} (\bibinfo {year} {2014})}\BibitemShut {NoStop}%
\bibitem [{\citenamefont {Romano}\ \emph {et~al.}(2010)\citenamefont {Romano}, \citenamefont {Sanz},\ and\ \citenamefont {Sciortino}}]{Romano2010}%
  \BibitemOpen
  \bibfield  {author} {\bibinfo {author} {\bibfnamefont {F.}~\bibnamefont {Romano}}, \bibinfo {author} {\bibfnamefont {E.}~\bibnamefont {Sanz}}, \ and\ \bibinfo {author} {\bibfnamefont {F.}~\bibnamefont {Sciortino}},\ }\href@noop {} {\bibfield  {journal} {\bibinfo  {journal} {J. Chem. Phys.}\ }\textbf {\bibinfo {volume} {132}},\ \bibinfo {pages} {184501} (\bibinfo {year} {2010})}\BibitemShut {NoStop}%
\bibitem [{\citenamefont {Howard}\ \emph {et~al.}(2019)\citenamefont {Howard}, \citenamefont {Jadrich}, \citenamefont {Lindquist}, \citenamefont {Khabaz}, \citenamefont {Bonnecaze}, \citenamefont {Milliron},\ and\ \citenamefont {Truskett}}]{Howard2019}%
  \BibitemOpen
  \bibfield  {author} {\bibinfo {author} {\bibfnamefont {M.~P.}\ \bibnamefont {Howard}}, \bibinfo {author} {\bibfnamefont {R.~B.}\ \bibnamefont {Jadrich}}, \bibinfo {author} {\bibfnamefont {B.~A.}\ \bibnamefont {Lindquist}}, \bibinfo {author} {\bibfnamefont {F.}~\bibnamefont {Khabaz}}, \bibinfo {author} {\bibfnamefont {R.~T.}\ \bibnamefont {Bonnecaze}}, \bibinfo {author} {\bibfnamefont {D.~J.}\ \bibnamefont {Milliron}}, \ and\ \bibinfo {author} {\bibfnamefont {T.~M.}\ \bibnamefont {Truskett}},\ }\href@noop {} {\bibfield  {journal} {\bibinfo  {journal} {J. Chem. Phys.}\ }\textbf {\bibinfo {volume} {151}},\ \bibinfo {pages} {124901} (\bibinfo {year} {2019})}\BibitemShut {NoStop}%
\bibitem [{\citenamefont {Zaccarelli}\ \emph {et~al.}(2006)\citenamefont {Zaccarelli}, \citenamefont {Saika-Voivod}, \citenamefont {Buldyrev}, \citenamefont {Moreno}, \citenamefont {Tartaglia},\ and\ \citenamefont {Sciortino}}]{Zaccarelli2006}%
  \BibitemOpen
  \bibfield  {author} {\bibinfo {author} {\bibfnamefont {E.}~\bibnamefont {Zaccarelli}}, \bibinfo {author} {\bibfnamefont {I.}~\bibnamefont {Saika-Voivod}}, \bibinfo {author} {\bibfnamefont {S.~V.}\ \bibnamefont {Buldyrev}}, \bibinfo {author} {\bibfnamefont {A.~J.}\ \bibnamefont {Moreno}}, \bibinfo {author} {\bibfnamefont {P.}~\bibnamefont {Tartaglia}}, \ and\ \bibinfo {author} {\bibfnamefont {F.}~\bibnamefont {Sciortino}},\ }\href@noop {} {\bibfield  {journal} {\bibinfo  {journal} {J. Chem. Phys.}\ }\textbf {\bibinfo {volume} {124}},\ \bibinfo {pages} {124908} (\bibinfo {year} {2006})}\BibitemShut {NoStop}%
\bibitem [{\citenamefont {Sciortino}\ and\ \citenamefont {Zaccarelli}(2017)}]{Sciortino2017}%
  \BibitemOpen
  \bibfield  {author} {\bibinfo {author} {\bibfnamefont {F.}~\bibnamefont {Sciortino}}\ and\ \bibinfo {author} {\bibfnamefont {E.}~\bibnamefont {Zaccarelli}},\ }\href@noop {} {\bibfield  {journal} {\bibinfo  {journal} {Curr. Op. Coll. Interf. Sci.}\ }\textbf {\bibinfo {volume} {30}},\ \bibinfo {pages} {90} (\bibinfo {year} {2017})}\BibitemShut {NoStop}%
\bibitem [{\citenamefont {Capelot}\ \emph {et~al.}(2012)\citenamefont {Capelot}, \citenamefont {Unterlass}, \citenamefont {Tournilhac},\ and\ \citenamefont {Leibler}}]{Capelot2012}%
  \BibitemOpen
  \bibfield  {author} {\bibinfo {author} {\bibfnamefont {M.}~\bibnamefont {Capelot}}, \bibinfo {author} {\bibfnamefont {M.~M.}\ \bibnamefont {Unterlass}}, \bibinfo {author} {\bibfnamefont {F.}~\bibnamefont {Tournilhac}}, \ and\ \bibinfo {author} {\bibfnamefont {L.}~\bibnamefont {Leibler}},\ }\href@noop {} {\bibfield  {journal} {\bibinfo  {journal} {ACS Macro Lett.}\ }\textbf {\bibinfo {volume} {1}},\ \bibinfo {pages} {789} (\bibinfo {year} {2012})}\BibitemShut {NoStop}%
\bibitem [{\citenamefont {Zaccarelli}\ \emph {et~al.}(2005)\citenamefont {Zaccarelli}, \citenamefont {Buldyrev}, \citenamefont {Nave}, \citenamefont {Moreno}, \citenamefont {Saika-Voivod}, \citenamefont {Sciortino},\ and\ \citenamefont {Tartaglia}}]{Zaccarelli2005}%
  \BibitemOpen
  \bibfield  {author} {\bibinfo {author} {\bibfnamefont {E.}~\bibnamefont {Zaccarelli}}, \bibinfo {author} {\bibfnamefont {S.~V.}\ \bibnamefont {Buldyrev}}, \bibinfo {author} {\bibfnamefont {E.~L.}\ \bibnamefont {Nave}}, \bibinfo {author} {\bibfnamefont {A.~J.}\ \bibnamefont {Moreno}}, \bibinfo {author} {\bibfnamefont {I.}~\bibnamefont {Saika-Voivod}}, \bibinfo {author} {\bibfnamefont {F.}~\bibnamefont {Sciortino}}, \ and\ \bibinfo {author} {\bibfnamefont {P.}~\bibnamefont {Tartaglia}},\ }\href@noop {} {\bibfield  {journal} {\bibinfo  {journal} {Phys. Rev. Lett.}\ }\textbf {\bibinfo {volume} {94}},\ \bibinfo {pages} {218301} (\bibinfo {year} {2005})}\BibitemShut {NoStop}%
\bibitem [{\citenamefont {Kim}\ \emph {et~al.}(2023)\citenamefont {Kim}, \citenamefont {Moon}, \citenamefont {Hwang}, \citenamefont {Kim},\ and\ \citenamefont {Yi}}]{Kim2023}%
  \BibitemOpen
  \bibfield  {author} {\bibinfo {author} {\bibfnamefont {Y.~J.}\ \bibnamefont {Kim}}, \bibinfo {author} {\bibfnamefont {J.~B.}\ \bibnamefont {Moon}}, \bibinfo {author} {\bibfnamefont {H.}~\bibnamefont {Hwang}}, \bibinfo {author} {\bibfnamefont {Y.~S.}\ \bibnamefont {Kim}}, \ and\ \bibinfo {author} {\bibfnamefont {G.~R.}\ \bibnamefont {Yi}},\ }\href@noop {} {\bibfield  {journal} {\bibinfo  {journal} {Adv. Mater.}\ }\textbf {\bibinfo {volume} {35}},\ \bibinfo {pages} {2203045} (\bibinfo {year} {2023})}\BibitemShut {NoStop}%
\bibitem [{\citenamefont {Wang}\ \emph {et~al.}(2012)\citenamefont {Wang}, \citenamefont {Wang}, \citenamefont {Breed}, \citenamefont {Manoharan}, \citenamefont {Feng}, \citenamefont {Hollingsworth}, \citenamefont {Weck},\ and\ \citenamefont {Pine}}]{Wang2012}%
  \BibitemOpen
  \bibfield  {author} {\bibinfo {author} {\bibfnamefont {Y.}~\bibnamefont {Wang}}, \bibinfo {author} {\bibfnamefont {Y.}~\bibnamefont {Wang}}, \bibinfo {author} {\bibfnamefont {D.~R.}\ \bibnamefont {Breed}}, \bibinfo {author} {\bibfnamefont {V.~N.}\ \bibnamefont {Manoharan}}, \bibinfo {author} {\bibfnamefont {L.}~\bibnamefont {Feng}}, \bibinfo {author} {\bibfnamefont {A.~D.}\ \bibnamefont {Hollingsworth}}, \bibinfo {author} {\bibfnamefont {M.}~\bibnamefont {Weck}}, \ and\ \bibinfo {author} {\bibfnamefont {D.~J.}\ \bibnamefont {Pine}},\ }\href@noop {} {\bibfield  {journal} {\bibinfo  {journal} {Nature}\ }\textbf {\bibinfo {volume} {491}},\ \bibinfo {pages} {51} (\bibinfo {year} {2012})}\BibitemShut {NoStop}%
\bibitem [{\citenamefont {Russo}\ \emph {et~al.}(2009)\citenamefont {Russo}, \citenamefont {Tartaglia},\ and\ \citenamefont {Sciortino}}]{Russo2009}%
  \BibitemOpen
  \bibfield  {author} {\bibinfo {author} {\bibfnamefont {J.}~\bibnamefont {Russo}}, \bibinfo {author} {\bibfnamefont {P.}~\bibnamefont {Tartaglia}}, \ and\ \bibinfo {author} {\bibfnamefont {F.}~\bibnamefont {Sciortino}},\ }\href@noop {} {\bibfield  {journal} {\bibinfo  {journal} {J. Chem. Phys.}\ }\textbf {\bibinfo {volume} {131}},\ \bibinfo {pages} {014504} (\bibinfo {year} {2009})}\BibitemShut {NoStop}%
\bibitem [{\citenamefont {Smallenburg}\ and\ \citenamefont {Sciortino}(2013)}]{Smallenburg2013}%
  \BibitemOpen
  \bibfield  {author} {\bibinfo {author} {\bibfnamefont {F.}~\bibnamefont {Smallenburg}}\ and\ \bibinfo {author} {\bibfnamefont {F.}~\bibnamefont {Sciortino}},\ }\href@noop {} {\bibfield  {journal} {\bibinfo  {journal} {Nature Phys.}\ }\textbf {\bibinfo {volume} {9}},\ \bibinfo {pages} {554} (\bibinfo {year} {2013})}\BibitemShut {NoStop}%
\bibitem [{\citenamefont {Li}\ \emph {et~al.}(2020)\citenamefont {Li}, \citenamefont {Palis}, \citenamefont {Mérindol}, \citenamefont {Majimel}, \citenamefont {Ravaine},\ and\ \citenamefont {Duguet}}]{Li2020}%
  \BibitemOpen
  \bibfield  {author} {\bibinfo {author} {\bibfnamefont {W.}~\bibnamefont {Li}}, \bibinfo {author} {\bibfnamefont {H.}~\bibnamefont {Palis}}, \bibinfo {author} {\bibfnamefont {R.}~\bibnamefont {Mérindol}}, \bibinfo {author} {\bibfnamefont {J.}~\bibnamefont {Majimel}}, \bibinfo {author} {\bibfnamefont {S.}~\bibnamefont {Ravaine}}, \ and\ \bibinfo {author} {\bibfnamefont {E.}~\bibnamefont {Duguet}},\ }\href@noop {} {\bibfield  {journal} {\bibinfo  {journal} {Chem. Soc. Rev.}\ }\textbf {\bibinfo {volume} {49}},\ \bibinfo {pages} {1955} (\bibinfo {year} {2020})}\BibitemShut {NoStop}%
\bibitem [{\citenamefont {Dias}\ \emph {et~al.}(2013{\natexlab{a}})\citenamefont {Dias}, \citenamefont {Araújo},\ and\ \citenamefont {da~Gama}}]{Dias2013b}%
  \BibitemOpen
  \bibfield  {author} {\bibinfo {author} {\bibfnamefont {C.~S.}\ \bibnamefont {Dias}}, \bibinfo {author} {\bibfnamefont {N.~A.~M.}\ \bibnamefont {Araújo}}, \ and\ \bibinfo {author} {\bibfnamefont {M.~M.~T.}\ \bibnamefont {da~Gama}},\ }\href@noop {} {\bibfield  {journal} {\bibinfo  {journal} {J. Chem. Phys.}\ }\textbf {\bibinfo {volume} {139}},\ \bibinfo {pages} {154903} (\bibinfo {year} {2013}{\natexlab{a}})}\BibitemShut {NoStop}%
\bibitem [{\citenamefont {de~las Heras}\ \emph {et~al.}(2011{\natexlab{a}})\citenamefont {de~las Heras}, \citenamefont {Tavares},\ and\ \citenamefont {da~Gama}}]{DelasHeras2010}%
  \BibitemOpen
  \bibfield  {author} {\bibinfo {author} {\bibfnamefont {D.}~\bibnamefont {de~las Heras}}, \bibinfo {author} {\bibfnamefont {J.~M.}\ \bibnamefont {Tavares}}, \ and\ \bibinfo {author} {\bibfnamefont {M.~M.~T.}\ \bibnamefont {da~Gama}},\ }\href@noop {} {\bibfield  {journal} {\bibinfo  {journal} {J. Chem. Phys.}\ }\textbf {\bibinfo {volume} {134}},\ \bibinfo {pages} {104904} (\bibinfo {year} {2011}{\natexlab{a}})}\BibitemShut {NoStop}%
\bibitem [{\citenamefont {Słyk}\ \emph {et~al.}(2016)\citenamefont {Słyk}, \citenamefont {Rżysko},\ and\ \citenamefont {Bryk}}]{Syk2016}%
  \BibitemOpen
  \bibfield  {author} {\bibinfo {author} {\bibfnamefont {E.}~\bibnamefont {Słyk}}, \bibinfo {author} {\bibfnamefont {W.}~\bibnamefont {Rżysko}}, \ and\ \bibinfo {author} {\bibfnamefont {P.}~\bibnamefont {Bryk}},\ }\href@noop {} {\bibfield  {journal} {\bibinfo  {journal} {Soft Matt.}\ }\textbf {\bibinfo {volume} {12}},\ \bibinfo {pages} {9538} (\bibinfo {year} {2016})}\BibitemShut {NoStop}%
\bibitem [{\citenamefont {Dias}\ \emph {et~al.}(2013{\natexlab{b}})\citenamefont {Dias}, \citenamefont {Araújo},\ and\ \citenamefont {da~Gama}}]{Dias2013}%
  \BibitemOpen
  \bibfield  {author} {\bibinfo {author} {\bibfnamefont {C.~S.}\ \bibnamefont {Dias}}, \bibinfo {author} {\bibfnamefont {N.~A.~M.}\ \bibnamefont {Araújo}}, \ and\ \bibinfo {author} {\bibfnamefont {M.~M.~T.}\ \bibnamefont {da~Gama}},\ }\href@noop {} {\bibfield  {journal} {\bibinfo  {journal} {Phys. Rev. E}\ }\textbf {\bibinfo {volume} {87}},\ \bibinfo {pages} {032308} (\bibinfo {year} {2013}{\natexlab{b}})}\BibitemShut {NoStop}%
\bibitem [{\citenamefont {Swinkels}\ \emph {et~al.}(2024)\citenamefont {Swinkels}, \citenamefont {Sinaasappel}, \citenamefont {Gong}, \citenamefont {Sacanna}, \citenamefont {Meyer}, \citenamefont {Sciortino},\ and\ \citenamefont {Schall}}]{Swinkels2024}%
  \BibitemOpen
  \bibfield  {author} {\bibinfo {author} {\bibfnamefont {P.~J.}\ \bibnamefont {Swinkels}}, \bibinfo {author} {\bibfnamefont {R.}~\bibnamefont {Sinaasappel}}, \bibinfo {author} {\bibfnamefont {Z.}~\bibnamefont {Gong}}, \bibinfo {author} {\bibfnamefont {S.}~\bibnamefont {Sacanna}}, \bibinfo {author} {\bibfnamefont {W.~V.}\ \bibnamefont {Meyer}}, \bibinfo {author} {\bibfnamefont {F.}~\bibnamefont {Sciortino}}, \ and\ \bibinfo {author} {\bibfnamefont {P.}~\bibnamefont {Schall}},\ }\href@noop {} {\bibfield  {journal} {\bibinfo  {journal} {Phys. Rev. Lett.}\ }\textbf {\bibinfo {volume} {132}},\ \bibinfo {pages} {078203} (\bibinfo {year} {2024})}\BibitemShut {NoStop}%
\bibitem [{\citenamefont {Liu}\ \emph {et~al.}(2020)\citenamefont {Liu}, \citenamefont {Zheng}, \citenamefont {Grebe}, \citenamefont {Pine},\ and\ \citenamefont {Weck}}]{Liu2020a}%
  \BibitemOpen
  \bibfield  {author} {\bibinfo {author} {\bibfnamefont {M.}~\bibnamefont {Liu}}, \bibinfo {author} {\bibfnamefont {X.}~\bibnamefont {Zheng}}, \bibinfo {author} {\bibfnamefont {V.}~\bibnamefont {Grebe}}, \bibinfo {author} {\bibfnamefont {D.~J.}\ \bibnamefont {Pine}}, \ and\ \bibinfo {author} {\bibfnamefont {M.}~\bibnamefont {Weck}},\ }\href@noop {} {\bibfield  {journal} {\bibinfo  {journal} {Nat. Mater.}\ }\textbf {\bibinfo {volume} {19}},\ \bibinfo {pages} {1354} (\bibinfo {year} {2020})}\BibitemShut {NoStop}%
\bibitem [{\citenamefont {de~las Heras}\ and\ \citenamefont {Schmidt}(2013)}]{DelasHeras2013}%
  \BibitemOpen
  \bibfield  {author} {\bibinfo {author} {\bibfnamefont {D.}~\bibnamefont {de~las Heras}}\ and\ \bibinfo {author} {\bibfnamefont {M.}~\bibnamefont {Schmidt}},\ }\href@noop {} {\bibfield  {journal} {\bibinfo  {journal} {Soft Matt.}\ }\textbf {\bibinfo {volume} {9}},\ \bibinfo {pages} {8636} (\bibinfo {year} {2013})}\BibitemShut {NoStop}%
\bibitem [{\citenamefont {Teixeira}\ \emph {et~al.}(2021)\citenamefont {Teixeira}, \citenamefont {de~las Heras}, \citenamefont {Tavares},\ and\ \citenamefont {da~Gama}}]{Teixeira2021}%
  \BibitemOpen
  \bibfield  {author} {\bibinfo {author} {\bibfnamefont {R.~B.}\ \bibnamefont {Teixeira}}, \bibinfo {author} {\bibfnamefont {D.}~\bibnamefont {de~las Heras}}, \bibinfo {author} {\bibfnamefont {J.~M.}\ \bibnamefont {Tavares}}, \ and\ \bibinfo {author} {\bibfnamefont {M.~M.~T.}\ \bibnamefont {da~Gama}},\ }\href@noop {} {\bibfield  {journal} {\bibinfo  {journal} {J. Chem. Phys.}\ }\textbf {\bibinfo {volume} {155}},\ \bibinfo {pages} {044903} (\bibinfo {year} {2021})}\BibitemShut {NoStop}%
\bibitem [{\citenamefont {Geigenfeind}\ and\ \citenamefont {Heras}(2017)}]{Geigenfeind2017}%
  \BibitemOpen
  \bibfield  {author} {\bibinfo {author} {\bibfnamefont {T.}~\bibnamefont {Geigenfeind}}\ and\ \bibinfo {author} {\bibfnamefont {D.~D.~L.}\ \bibnamefont {Heras}},\ }\href@noop {} {\bibfield  {journal} {\bibinfo  {journal} {J. Phys.: Cond. Matt.}\ }\textbf {\bibinfo {volume} {29}},\ \bibinfo {pages} {064006} (\bibinfo {year} {2017})}\BibitemShut {NoStop}%
\bibitem [{\citenamefont {Araújo}\ \emph {et~al.}(2015)\citenamefont {Araújo}, \citenamefont {Dias},\ and\ \citenamefont {da~Gama}}]{Araujo2015}%
  \BibitemOpen
  \bibfield  {author} {\bibinfo {author} {\bibfnamefont {N.~A.~M.}\ \bibnamefont {Araújo}}, \bibinfo {author} {\bibfnamefont {C.~S.}\ \bibnamefont {Dias}}, \ and\ \bibinfo {author} {\bibfnamefont {M.~M.~T.}\ \bibnamefont {da~Gama}},\ }\href@noop {} {\bibfield  {journal} {\bibinfo  {journal} {J. Phys.: Condens. Matter}\ }\textbf {\bibinfo {volume} {27}},\ \bibinfo {pages} {194123} (\bibinfo {year} {2015})}\BibitemShut {NoStop}%
\bibitem [{\citenamefont {Dias}\ \emph {et~al.}(2016)\citenamefont {Dias}, \citenamefont {Braga}, \citenamefont {Araújo},\ and\ \citenamefont {da~Gama}}]{Dias2016}%
  \BibitemOpen
  \bibfield  {author} {\bibinfo {author} {\bibfnamefont {C.~S.}\ \bibnamefont {Dias}}, \bibinfo {author} {\bibfnamefont {C.}~\bibnamefont {Braga}}, \bibinfo {author} {\bibfnamefont {N.~A.~M.}\ \bibnamefont {Araújo}}, \ and\ \bibinfo {author} {\bibfnamefont {M.~M.~T.}\ \bibnamefont {da~Gama}},\ }\href@noop {} {\bibfield  {journal} {\bibinfo  {journal} {Soft Matt.}\ }\textbf {\bibinfo {volume} {12}},\ \bibinfo {pages} {1550} (\bibinfo {year} {2016})}\BibitemShut {NoStop}%
\bibitem [{\citenamefont {Araújo}\ \emph {et~al.}(2017)\citenamefont {Araújo}, \citenamefont {Dias},\ and\ \citenamefont {da~Gama}}]{Araujo2017}%
  \BibitemOpen
  \bibfield  {author} {\bibinfo {author} {\bibfnamefont {N.~A.~M.}\ \bibnamefont {Araújo}}, \bibinfo {author} {\bibfnamefont {C.~S.}\ \bibnamefont {Dias}}, \ and\ \bibinfo {author} {\bibfnamefont {M.~M.~T.}\ \bibnamefont {da~Gama}},\ }\href@noop {} {\bibfield  {journal} {\bibinfo  {journal} {J. Phys.: Condens. Matter}\ }\textbf {\bibinfo {volume} {29}},\ \bibinfo {pages} {014001} (\bibinfo {year} {2017})}\BibitemShut {NoStop}%
\bibitem [{\citenamefont {Dias}\ \emph {et~al.}(2015)\citenamefont {Dias}, \citenamefont {Araújo},\ and\ \citenamefont {da~Gama}}]{Dias2015}%
  \BibitemOpen
  \bibfield  {author} {\bibinfo {author} {\bibfnamefont {C.~S.}\ \bibnamefont {Dias}}, \bibinfo {author} {\bibfnamefont {N.~A.~M.}\ \bibnamefont {Araújo}}, \ and\ \bibinfo {author} {\bibfnamefont {M.~M.~T.}\ \bibnamefont {da~Gama}},\ }\href@noop {} {\bibfield  {journal} {\bibinfo  {journal} {Mol. Phys.}\ }\textbf {\bibinfo {volume} {113}},\ \bibinfo {pages} {1069} (\bibinfo {year} {2015})}\BibitemShut {NoStop}%
\bibitem [{\citenamefont {Dias}\ \emph {et~al.}(2018{\natexlab{a}})\citenamefont {Dias}, \citenamefont {Yunker}, \citenamefont {Yodh}, \citenamefont {Araújo},\ and\ \citenamefont {da~Gama}}]{Dias2018a}%
  \BibitemOpen
  \bibfield  {author} {\bibinfo {author} {\bibfnamefont {C.~S.}\ \bibnamefont {Dias}}, \bibinfo {author} {\bibfnamefont {P.~J.}\ \bibnamefont {Yunker}}, \bibinfo {author} {\bibfnamefont {A.~G.}\ \bibnamefont {Yodh}}, \bibinfo {author} {\bibfnamefont {N.~A.~M.}\ \bibnamefont {Araújo}}, \ and\ \bibinfo {author} {\bibfnamefont {M.~M.~T.}\ \bibnamefont {da~Gama}},\ }\href@noop {} {\bibfield  {journal} {\bibinfo  {journal} {Soft Matt.}\ }\textbf {\bibinfo {volume} {14}},\ \bibinfo {pages} {1903} (\bibinfo {year} {2018}{\natexlab{a}})}\BibitemShut {NoStop}%
\bibitem [{\citenamefont {Fenton}\ \emph {et~al.}(2018)\citenamefont {Fenton}, \citenamefont {Steimle},\ and\ \citenamefont {Schaak}}]{Fenton2018}%
  \BibitemOpen
  \bibfield  {author} {\bibinfo {author} {\bibfnamefont {J.~L.}\ \bibnamefont {Fenton}}, \bibinfo {author} {\bibfnamefont {B.~C.}\ \bibnamefont {Steimle}}, \ and\ \bibinfo {author} {\bibfnamefont {R.~E.}\ \bibnamefont {Schaak}},\ }\href@noop {} {\bibfield  {journal} {\bibinfo  {journal} {Science}\ }\textbf {\bibinfo {volume} {360}},\ \bibinfo {pages} {513} (\bibinfo {year} {2018})}\BibitemShut {NoStop}%
\bibitem [{\citenamefont {Dias}\ \emph {et~al.}(2014{\natexlab{a}})\citenamefont {Dias}, \citenamefont {Araújo},\ and\ \citenamefont {da~Gama}}]{Dias2014}%
  \BibitemOpen
  \bibfield  {author} {\bibinfo {author} {\bibfnamefont {C.~S.}\ \bibnamefont {Dias}}, \bibinfo {author} {\bibfnamefont {N.~A.~M.}\ \bibnamefont {Araújo}}, \ and\ \bibinfo {author} {\bibfnamefont {M.~M.~T.}\ \bibnamefont {da~Gama}},\ }\href@noop {} {\bibfield  {journal} {\bibinfo  {journal} {Phys. Rev. E}\ }\textbf {\bibinfo {volume} {90}},\ \bibinfo {pages} {032302} (\bibinfo {year} {2014}{\natexlab{a}})}\BibitemShut {NoStop}%
\bibitem [{\citenamefont {Dias}\ \emph {et~al.}(2014{\natexlab{b}})\citenamefont {Dias}, \citenamefont {Araújo},\ and\ \citenamefont {da~Gama}}]{Dias2014a}%
  \BibitemOpen
  \bibfield  {author} {\bibinfo {author} {\bibfnamefont {C.~S.}\ \bibnamefont {Dias}}, \bibinfo {author} {\bibfnamefont {N.~A.~M.}\ \bibnamefont {Araújo}}, \ and\ \bibinfo {author} {\bibfnamefont {M.~M.~T.}\ \bibnamefont {da~Gama}},\ }\href@noop {} {\bibfield  {journal} {\bibinfo  {journal} {EPL}\ }\textbf {\bibinfo {volume} {107}},\ \bibinfo {pages} {56002} (\bibinfo {year} {2014}{\natexlab{b}})}\BibitemShut {NoStop}%
\bibitem [{\citenamefont {Tavares}\ \emph {et~al.}(2020)\citenamefont {Tavares}, \citenamefont {Antunes}, \citenamefont {Dias}, \citenamefont {Gama},\ and\ \citenamefont {Araújo}}]{Tavares2020}%
  \BibitemOpen
  \bibfield  {author} {\bibinfo {author} {\bibfnamefont {J.~M.}\ \bibnamefont {Tavares}}, \bibinfo {author} {\bibfnamefont {G.~C.}\ \bibnamefont {Antunes}}, \bibinfo {author} {\bibfnamefont {C.~S.}\ \bibnamefont {Dias}}, \bibinfo {author} {\bibfnamefont {M.~M. T.~D.}\ \bibnamefont {Gama}}, \ and\ \bibinfo {author} {\bibfnamefont {N.~A.}\ \bibnamefont {Araújo}},\ }\href@noop {} {\bibfield  {journal} {\bibinfo  {journal} {Soft Matt.}\ }\textbf {\bibinfo {volume} {16}},\ \bibinfo {pages} {7513} (\bibinfo {year} {2020})}\BibitemShut {NoStop}%
\bibitem [{\citenamefont {Lei}\ \emph {et~al.}(2020)\citenamefont {Lei}, \citenamefont {Xia}, \citenamefont {Yang}, \citenamefont {Ciamarra},\ and\ \citenamefont {Ni}}]{Lei2020}%
  \BibitemOpen
  \bibfield  {author} {\bibinfo {author} {\bibfnamefont {Q.~L.}\ \bibnamefont {Lei}}, \bibinfo {author} {\bibfnamefont {X.}~\bibnamefont {Xia}}, \bibinfo {author} {\bibfnamefont {J.}~\bibnamefont {Yang}}, \bibinfo {author} {\bibfnamefont {M.~P.}\ \bibnamefont {Ciamarra}}, \ and\ \bibinfo {author} {\bibfnamefont {R.}~\bibnamefont {Ni}},\ }\href@noop {} {\bibfield  {journal} {\bibinfo  {journal} {Proc. Natl. Acad. Sci. U.S.A.}\ }\textbf {\bibinfo {volume} {117}},\ \bibinfo {pages} {27111} (\bibinfo {year} {2020})}\BibitemShut {NoStop}%
\bibitem [{\citenamefont {Antunes}\ \emph {et~al.}(2019)\citenamefont {Antunes}, \citenamefont {Dias}, \citenamefont {Gama},\ and\ \citenamefont {Araújo}}]{Antunes2019}%
  \BibitemOpen
  \bibfield  {author} {\bibinfo {author} {\bibfnamefont {G.~C.}\ \bibnamefont {Antunes}}, \bibinfo {author} {\bibfnamefont {C.~S.}\ \bibnamefont {Dias}}, \bibinfo {author} {\bibfnamefont {M.~M. T.~D.}\ \bibnamefont {Gama}}, \ and\ \bibinfo {author} {\bibfnamefont {N.~A.~M.}\ \bibnamefont {Araújo}},\ }\href@noop {} {\bibfield  {journal} {\bibinfo  {journal} {Soft Matt.}\ }\textbf {\bibinfo {volume} {15}},\ \bibinfo {pages} {3712} (\bibinfo {year} {2019})}\BibitemShut {NoStop}%
\bibitem [{\citenamefont {Lowensohn}\ \emph {et~al.}(2019)\citenamefont {Lowensohn}, \citenamefont {Oyarzún}, \citenamefont {Paliza}, \citenamefont {Mognetti},\ and\ \citenamefont {Rogers}}]{Lowensohn2019}%
  \BibitemOpen
  \bibfield  {author} {\bibinfo {author} {\bibfnamefont {J.}~\bibnamefont {Lowensohn}}, \bibinfo {author} {\bibfnamefont {B.}~\bibnamefont {Oyarzún}}, \bibinfo {author} {\bibfnamefont {G.~N.}\ \bibnamefont {Paliza}}, \bibinfo {author} {\bibfnamefont {B.~M.}\ \bibnamefont {Mognetti}}, \ and\ \bibinfo {author} {\bibfnamefont {W.~B.}\ \bibnamefont {Rogers}},\ }\href@noop {} {\bibfield  {journal} {\bibinfo  {journal} {Phys. Rev. X}\ }\textbf {\bibinfo {volume} {9}},\ \bibinfo {pages} {41054} (\bibinfo {year} {2019})}\BibitemShut {NoStop}%
\bibitem [{\citenamefont {van~der Meulen}\ and\ \citenamefont {Leunissen}(2013)}]{VanDerMeulen2013}%
  \BibitemOpen
  \bibfield  {author} {\bibinfo {author} {\bibfnamefont {S.~A.~J.}\ \bibnamefont {van~der Meulen}}\ and\ \bibinfo {author} {\bibfnamefont {M.~E.}\ \bibnamefont {Leunissen}},\ }\href@noop {} {\bibfield  {journal} {\bibinfo  {journal} {J. Am. Chem. Soc.}\ }\textbf {\bibinfo {volume} {135}},\ \bibinfo {pages} {15129} (\bibinfo {year} {2013})}\BibitemShut {NoStop}%
\bibitem [{\citenamefont {Xia}\ \emph {et~al.}(2020)\citenamefont {Xia}, \citenamefont {Hu}, \citenamefont {Ciamarra},\ and\ \citenamefont {Ni}}]{Xia2020}%
  \BibitemOpen
  \bibfield  {author} {\bibinfo {author} {\bibfnamefont {X.}~\bibnamefont {Xia}}, \bibinfo {author} {\bibfnamefont {H.}~\bibnamefont {Hu}}, \bibinfo {author} {\bibfnamefont {M.~P.}\ \bibnamefont {Ciamarra}}, \ and\ \bibinfo {author} {\bibfnamefont {R.}~\bibnamefont {Ni}},\ }\href@noop {} {\bibfield  {journal} {\bibinfo  {journal} {Sci. Adv.}\ }\textbf {\bibinfo {volume} {6}},\ \bibinfo {pages} {eaaz6921} (\bibinfo {year} {2020})}\BibitemShut {NoStop}%
\bibitem [{\citenamefont {Angioletti-Uberti}\ \emph {et~al.}(2014)\citenamefont {Angioletti-Uberti}, \citenamefont {Varilly}, \citenamefont {Mognetti},\ and\ \citenamefont {Frenkel}}]{Angioletti-Uberti2014}%
  \BibitemOpen
  \bibfield  {author} {\bibinfo {author} {\bibfnamefont {S.}~\bibnamefont {Angioletti-Uberti}}, \bibinfo {author} {\bibfnamefont {P.}~\bibnamefont {Varilly}}, \bibinfo {author} {\bibfnamefont {B.~M.}\ \bibnamefont {Mognetti}}, \ and\ \bibinfo {author} {\bibfnamefont {D.}~\bibnamefont {Frenkel}},\ }\href@noop {} {\bibfield  {journal} {\bibinfo  {journal} {Phys. Rev. Lett.}\ }\textbf {\bibinfo {volume} {113}},\ \bibinfo {pages} {128303} (\bibinfo {year} {2014})}\BibitemShut {NoStop}%
\bibitem [{\citenamefont {Xiong}\ \emph {et~al.}(2009)\citenamefont {Xiong}, \citenamefont {Lelie},\ and\ \citenamefont {Gang}}]{Xiong2009}%
  \BibitemOpen
  \bibfield  {author} {\bibinfo {author} {\bibfnamefont {H.}~\bibnamefont {Xiong}}, \bibinfo {author} {\bibfnamefont {D.~V.~D.}\ \bibnamefont {Lelie}}, \ and\ \bibinfo {author} {\bibfnamefont {O.}~\bibnamefont {Gang}},\ }\href@noop {} {\bibfield  {journal} {\bibinfo  {journal} {Phys. Rev. Lett.}\ }\textbf {\bibinfo {volume} {102}},\ \bibinfo {pages} {015504} (\bibinfo {year} {2009})}\BibitemShut {NoStop}%
\bibitem [{\citenamefont {Dias}\ \emph {et~al.}(2020)\citenamefont {Dias}, \citenamefont {Custodio}, \citenamefont {Antunes}, \citenamefont {da~Gama}, \citenamefont {Mano},\ and\ \citenamefont {Araujo}}]{Dias2020a}%
  \BibitemOpen
  \bibfield  {author} {\bibinfo {author} {\bibfnamefont {C.~S.}\ \bibnamefont {Dias}}, \bibinfo {author} {\bibfnamefont {C.~A.}\ \bibnamefont {Custodio}}, \bibinfo {author} {\bibfnamefont {G.~C.}\ \bibnamefont {Antunes}}, \bibinfo {author} {\bibfnamefont {M.~M.~T.}\ \bibnamefont {da~Gama}}, \bibinfo {author} {\bibfnamefont {J.~F.}\ \bibnamefont {Mano}}, \ and\ \bibinfo {author} {\bibfnamefont {N.~A.}\ \bibnamefont {Araujo}},\ }\href@noop {} {\bibfield  {journal} {\bibinfo  {journal} {ACS Appl. Mater. Interfaces}\ }\textbf {\bibinfo {volume} {12}},\ \bibinfo {pages} {48321} (\bibinfo {year} {2020})}\BibitemShut {NoStop}%
\bibitem [{\citenamefont {Kang}\ \emph {et~al.}(2023)\citenamefont {Kang}, \citenamefont {Sherman}, \citenamefont {Conrad}, \citenamefont {Crory}, \citenamefont {Dominguez}, \citenamefont {Valenzuela}, \citenamefont {Anslyn}, \citenamefont {Truskett},\ and\ \citenamefont {Milliron}}]{Kang2023}%
  \BibitemOpen
  \bibfield  {author} {\bibinfo {author} {\bibfnamefont {J.}~\bibnamefont {Kang}}, \bibinfo {author} {\bibfnamefont {Z.~M.}\ \bibnamefont {Sherman}}, \bibinfo {author} {\bibfnamefont {D.~L.}\ \bibnamefont {Conrad}}, \bibinfo {author} {\bibfnamefont {H.~S.~N.}\ \bibnamefont {Crory}}, \bibinfo {author} {\bibfnamefont {M.~N.}\ \bibnamefont {Dominguez}}, \bibinfo {author} {\bibfnamefont {S.~A.}\ \bibnamefont {Valenzuela}}, \bibinfo {author} {\bibfnamefont {E.~V.}\ \bibnamefont {Anslyn}}, \bibinfo {author} {\bibfnamefont {T.~M.}\ \bibnamefont {Truskett}}, \ and\ \bibinfo {author} {\bibfnamefont {D.~J.}\ \bibnamefont {Milliron}},\ }\href@noop {} {\bibfield  {journal} {\bibinfo  {journal} {ACS Nano}\ }\textbf {\bibinfo {volume} {17}},\ \bibinfo {pages} {24218} (\bibinfo {year} {2023})}\BibitemShut {NoStop}%
\bibitem [{\citenamefont {Ruzicka}\ \emph {et~al.}(2011)\citenamefont {Ruzicka}, \citenamefont {Zaccarelli}, \citenamefont {Zulian}, \citenamefont {Angelini}, \citenamefont {Sztucki}, \citenamefont {Moussaïd}, \citenamefont {Narayanan},\ and\ \citenamefont {Sciortino}}]{Ruzicka2011}%
  \BibitemOpen
  \bibfield  {author} {\bibinfo {author} {\bibfnamefont {B.}~\bibnamefont {Ruzicka}}, \bibinfo {author} {\bibfnamefont {E.}~\bibnamefont {Zaccarelli}}, \bibinfo {author} {\bibfnamefont {L.}~\bibnamefont {Zulian}}, \bibinfo {author} {\bibfnamefont {R.}~\bibnamefont {Angelini}}, \bibinfo {author} {\bibfnamefont {M.}~\bibnamefont {Sztucki}}, \bibinfo {author} {\bibfnamefont {A.}~\bibnamefont {Moussaïd}}, \bibinfo {author} {\bibfnamefont {T.}~\bibnamefont {Narayanan}}, \ and\ \bibinfo {author} {\bibfnamefont {F.}~\bibnamefont {Sciortino}},\ }\href@noop {} {\bibfield  {journal} {\bibinfo  {journal} {Nat. Mater.}\ }\textbf {\bibinfo {volume} {10}},\ \bibinfo {pages} {56} (\bibinfo {year} {2011})}\BibitemShut {NoStop}%
\bibitem [{\citenamefont {Smallenburg}\ \emph {et~al.}(2013)\citenamefont {Smallenburg}, \citenamefont {Leibler},\ and\ \citenamefont {Sciortino}}]{Smallenburg2013a}%
  \BibitemOpen
  \bibfield  {author} {\bibinfo {author} {\bibfnamefont {F.}~\bibnamefont {Smallenburg}}, \bibinfo {author} {\bibfnamefont {L.}~\bibnamefont {Leibler}}, \ and\ \bibinfo {author} {\bibfnamefont {F.}~\bibnamefont {Sciortino}},\ }\href@noop {} {\bibfield  {journal} {\bibinfo  {journal} {Phys. Rev. Lett.}\ }\textbf {\bibinfo {volume} {111}},\ \bibinfo {pages} {188002} (\bibinfo {year} {2013})}\BibitemShut {NoStop}%
\bibitem [{\citenamefont {Sacanna}\ \emph {et~al.}(2013)\citenamefont {Sacanna}, \citenamefont {Korpics}, \citenamefont {Rodriguez}, \citenamefont {Colón-Meléndez}, \citenamefont {Kim}, \citenamefont {Pine},\ and\ \citenamefont {Yi}}]{Sacanna2013}%
  \BibitemOpen
  \bibfield  {author} {\bibinfo {author} {\bibfnamefont {S.}~\bibnamefont {Sacanna}}, \bibinfo {author} {\bibfnamefont {M.}~\bibnamefont {Korpics}}, \bibinfo {author} {\bibfnamefont {K.}~\bibnamefont {Rodriguez}}, \bibinfo {author} {\bibfnamefont {L.}~\bibnamefont {Colón-Meléndez}}, \bibinfo {author} {\bibfnamefont {S.-H.}\ \bibnamefont {Kim}}, \bibinfo {author} {\bibfnamefont {D.~J.}\ \bibnamefont {Pine}}, \ and\ \bibinfo {author} {\bibfnamefont {G.-R.}\ \bibnamefont {Yi}},\ }\href@noop {} {\bibfield  {journal} {\bibinfo  {journal} {Nat. Commun.}\ }\textbf {\bibinfo {volume} {4}},\ \bibinfo {pages} {1688} (\bibinfo {year} {2013})}\BibitemShut {NoStop}%
\bibitem [{\citenamefont {Shah}\ \emph {et~al.}(2013)\citenamefont {Shah}, \citenamefont {Schultz}, \citenamefont {Kohlstedt}, \citenamefont {Glotzer},\ and\ \citenamefont {Solomon}}]{Shah2013}%
  \BibitemOpen
  \bibfield  {author} {\bibinfo {author} {\bibfnamefont {A.~A.}\ \bibnamefont {Shah}}, \bibinfo {author} {\bibfnamefont {B.}~\bibnamefont {Schultz}}, \bibinfo {author} {\bibfnamefont {K.~L.}\ \bibnamefont {Kohlstedt}}, \bibinfo {author} {\bibfnamefont {S.~C.}\ \bibnamefont {Glotzer}}, \ and\ \bibinfo {author} {\bibfnamefont {M.~J.}\ \bibnamefont {Solomon}},\ }\href@noop {} {\bibfield  {journal} {\bibinfo  {journal} {Langmuir}\ }\textbf {\bibinfo {volume} {29}},\ \bibinfo {pages} {4688} (\bibinfo {year} {2013})}\BibitemShut {NoStop}%
\bibitem [{\citenamefont {van Blaaderen}\ \emph {et~al.}(2013)\citenamefont {van Blaaderen}, \citenamefont {Dijkstra}, \citenamefont {van Roij}, \citenamefont {Imhof}, \citenamefont {Kamp}, \citenamefont {Kwaadgras}, \citenamefont {Vissers},\ and\ \citenamefont {Liu}}]{VanBlaaderen2013}%
  \BibitemOpen
  \bibfield  {author} {\bibinfo {author} {\bibfnamefont {A.}~\bibnamefont {van Blaaderen}}, \bibinfo {author} {\bibfnamefont {M.}~\bibnamefont {Dijkstra}}, \bibinfo {author} {\bibfnamefont {R.}~\bibnamefont {van Roij}}, \bibinfo {author} {\bibfnamefont {A.}~\bibnamefont {Imhof}}, \bibinfo {author} {\bibfnamefont {M.}~\bibnamefont {Kamp}}, \bibinfo {author} {\bibfnamefont {B.~W.}\ \bibnamefont {Kwaadgras}}, \bibinfo {author} {\bibfnamefont {T.}~\bibnamefont {Vissers}}, \ and\ \bibinfo {author} {\bibfnamefont {B.}~\bibnamefont {Liu}},\ }\href@noop {} {\bibfield  {journal} {\bibinfo  {journal} {Euro. Phys. J. Spec. Top.}\ }\textbf {\bibinfo {volume} {222}},\ \bibinfo {pages} {2895} (\bibinfo {year} {2013})}\BibitemShut {NoStop}%
\bibitem [{\citenamefont {Dias}\ \emph {et~al.}(2017)\citenamefont {Dias}, \citenamefont {Araújo},\ and\ \citenamefont {da~Gama}}]{Dias2017}%
  \BibitemOpen
  \bibfield  {author} {\bibinfo {author} {\bibfnamefont {C.~S.}\ \bibnamefont {Dias}}, \bibinfo {author} {\bibfnamefont {N.~A.~M.}\ \bibnamefont {Araújo}}, \ and\ \bibinfo {author} {\bibfnamefont {M.~M.~T.}\ \bibnamefont {da~Gama}},\ }\href@noop {} {\bibfield  {journal} {\bibinfo  {journal} {Adv. Col. Interf. Sci.}\ }\textbf {\bibinfo {volume} {247}},\ \bibinfo {pages} {258} (\bibinfo {year} {2017})}\BibitemShut {NoStop}%
\bibitem [{\citenamefont {Ortiz}\ \emph {et~al.}(2014)\citenamefont {Ortiz}, \citenamefont {Kohlstedt}, \citenamefont {Nguyen},\ and\ \citenamefont {Glotzer}}]{Ortiz2014}%
  \BibitemOpen
  \bibfield  {author} {\bibinfo {author} {\bibfnamefont {D.}~\bibnamefont {Ortiz}}, \bibinfo {author} {\bibfnamefont {K.~L.}\ \bibnamefont {Kohlstedt}}, \bibinfo {author} {\bibfnamefont {T.~D.}\ \bibnamefont {Nguyen}}, \ and\ \bibinfo {author} {\bibfnamefont {S.~C.}\ \bibnamefont {Glotzer}},\ }\href@noop {} {\bibfield  {journal} {\bibinfo  {journal} {Soft Matt.}\ }\textbf {\bibinfo {volume} {10}},\ \bibinfo {pages} {3541} (\bibinfo {year} {2014})}\BibitemShut {NoStop}%
\bibitem [{\citenamefont {King}\ \emph {et~al.}(2024)\citenamefont {King}, \citenamefont {Du}, \citenamefont {Zhu}, \citenamefont {Schoenholz},\ and\ \citenamefont {Brenner}}]{King2024}%
  \BibitemOpen
  \bibfield  {author} {\bibinfo {author} {\bibfnamefont {E.~M.}\ \bibnamefont {King}}, \bibinfo {author} {\bibfnamefont {C.~X.}\ \bibnamefont {Du}}, \bibinfo {author} {\bibfnamefont {Q.-Z.}\ \bibnamefont {Zhu}}, \bibinfo {author} {\bibfnamefont {S.~S.}\ \bibnamefont {Schoenholz}}, \ and\ \bibinfo {author} {\bibfnamefont {M.~P.}\ \bibnamefont {Brenner}},\ }\href@noop {} {\bibfield  {journal} {\bibinfo  {journal} {PNAS}\ }\textbf {\bibinfo {volume} {121}},\ \bibinfo {pages} {e2311891121} (\bibinfo {year} {2024})}\BibitemShut {NoStop}%
\bibitem [{\citenamefont {Nag}\ and\ \citenamefont {Bisker}(2024)}]{Nag2024}%
  \BibitemOpen
  \bibfield  {author} {\bibinfo {author} {\bibfnamefont {S.}~\bibnamefont {Nag}}\ and\ \bibinfo {author} {\bibfnamefont {G.}~\bibnamefont {Bisker}},\ }\href@noop {} {\bibfield  {journal} {\bibinfo  {journal} {J. Chem. Theory Comput}\ }\textbf {\bibinfo {volume} {20}},\ \bibinfo {pages} {8844} (\bibinfo {year} {2024})}\BibitemShut {NoStop}%
\bibitem [{\citenamefont {Dias}\ \emph {et~al.}(2018{\natexlab{b}})\citenamefont {Dias}, \citenamefont {Tavares}, \citenamefont {Araújo},\ and\ \citenamefont {da~Gama}}]{Dias2018b}%
  \BibitemOpen
  \bibfield  {author} {\bibinfo {author} {\bibfnamefont {C.~S.}\ \bibnamefont {Dias}}, \bibinfo {author} {\bibfnamefont {J.~M.}\ \bibnamefont {Tavares}}, \bibinfo {author} {\bibfnamefont {N.~A.~M.}\ \bibnamefont {Araújo}}, \ and\ \bibinfo {author} {\bibfnamefont {M.~M.~T.}\ \bibnamefont {da~Gama}},\ }\href@noop {} {\bibfield  {journal} {\bibinfo  {journal} {Soft Matt.}\ }\textbf {\bibinfo {volume} {14}},\ \bibinfo {pages} {2744} (\bibinfo {year} {2018}{\natexlab{b}})}\BibitemShut {NoStop}%
\bibitem [{\citenamefont {Wolters}\ \emph {et~al.}(2015)\citenamefont {Wolters}, \citenamefont {Avvisati}, \citenamefont {Hagemans}, \citenamefont {Vissers}, \citenamefont {Kraft}, \citenamefont {Dijkstra},\ and\ \citenamefont {Kegel}}]{Wolters2015}%
  \BibitemOpen
  \bibfield  {author} {\bibinfo {author} {\bibfnamefont {J.~R.}\ \bibnamefont {Wolters}}, \bibinfo {author} {\bibfnamefont {G.}~\bibnamefont {Avvisati}}, \bibinfo {author} {\bibfnamefont {F.}~\bibnamefont {Hagemans}}, \bibinfo {author} {\bibfnamefont {T.}~\bibnamefont {Vissers}}, \bibinfo {author} {\bibfnamefont {D.~J.}\ \bibnamefont {Kraft}}, \bibinfo {author} {\bibfnamefont {M.}~\bibnamefont {Dijkstra}}, \ and\ \bibinfo {author} {\bibfnamefont {W.~K.}\ \bibnamefont {Kegel}},\ }\href@noop {} {\bibfield  {journal} {\bibinfo  {journal} {Soft Matt.}\ }\textbf {\bibinfo {volume} {11}},\ \bibinfo {pages} {1067} (\bibinfo {year} {2015})}\BibitemShut {NoStop}%
\bibitem [{\citenamefont {Russo}\ \emph {et~al.}(2022)\citenamefont {Russo}, \citenamefont {Leoni}, \citenamefont {Martelli},\ and\ \citenamefont {Sciortino}}]{Russo2022}%
  \BibitemOpen
  \bibfield  {author} {\bibinfo {author} {\bibfnamefont {J.}~\bibnamefont {Russo}}, \bibinfo {author} {\bibfnamefont {F.}~\bibnamefont {Leoni}}, \bibinfo {author} {\bibfnamefont {F.}~\bibnamefont {Martelli}}, \ and\ \bibinfo {author} {\bibfnamefont {F.}~\bibnamefont {Sciortino}},\ }\href@noop {} {\bibfield  {journal} {\bibinfo  {journal} {Rep. Prog. Phys.}\ }\textbf {\bibinfo {volume} {85}},\ \bibinfo {pages} {016601} (\bibinfo {year} {2022})}\BibitemShut {NoStop}%
\bibitem [{\citenamefont {Feng}\ \emph {et~al.}(2013)\citenamefont {Feng}, \citenamefont {Dreyfus}, \citenamefont {Sha}, \citenamefont {Seeman},\ and\ \citenamefont {Chaikin}}]{Feng2013}%
  \BibitemOpen
  \bibfield  {author} {\bibinfo {author} {\bibfnamefont {L.}~\bibnamefont {Feng}}, \bibinfo {author} {\bibfnamefont {R.}~\bibnamefont {Dreyfus}}, \bibinfo {author} {\bibfnamefont {R.}~\bibnamefont {Sha}}, \bibinfo {author} {\bibfnamefont {N.~C.}\ \bibnamefont {Seeman}}, \ and\ \bibinfo {author} {\bibfnamefont {P.~M.}\ \bibnamefont {Chaikin}},\ }\href@noop {} {\bibfield  {journal} {\bibinfo  {journal} {Adv. Mater.}\ }\textbf {\bibinfo {volume} {25}},\ \bibinfo {pages} {2779} (\bibinfo {year} {2013})}\BibitemShut {NoStop}%
\bibitem [{\citenamefont {Gong}\ \emph {et~al.}(2017)\citenamefont {Gong}, \citenamefont {Hueckel}, \citenamefont {Yi},\ and\ \citenamefont {Sacanna}}]{Gong2017}%
  \BibitemOpen
  \bibfield  {author} {\bibinfo {author} {\bibfnamefont {Z.}~\bibnamefont {Gong}}, \bibinfo {author} {\bibfnamefont {T.}~\bibnamefont {Hueckel}}, \bibinfo {author} {\bibfnamefont {G.~R.}\ \bibnamefont {Yi}}, \ and\ \bibinfo {author} {\bibfnamefont {S.}~\bibnamefont {Sacanna}},\ }\href@noop {} {\bibfield  {journal} {\bibinfo  {journal} {Nature}\ }\textbf {\bibinfo {volume} {550}},\ \bibinfo {pages} {234} (\bibinfo {year} {2017})}\BibitemShut {NoStop}%
\bibitem [{\citenamefont {Bantawa}\ \emph {et~al.}(2021)\citenamefont {Bantawa}, \citenamefont {Fontaine-Seiler}, \citenamefont {Olmsted},\ and\ \citenamefont {Gado}}]{Bantawa2021}%
  \BibitemOpen
  \bibfield  {author} {\bibinfo {author} {\bibfnamefont {M.}~\bibnamefont {Bantawa}}, \bibinfo {author} {\bibfnamefont {W.~A.}\ \bibnamefont {Fontaine-Seiler}}, \bibinfo {author} {\bibfnamefont {P.~D.}\ \bibnamefont {Olmsted}}, \ and\ \bibinfo {author} {\bibfnamefont {E.~D.}\ \bibnamefont {Gado}},\ }\href@noop {} {\bibfield  {journal} {\bibinfo  {journal} {J. Phys. : Condens. Matter}\ }\textbf {\bibinfo {volume} {33}},\ \bibinfo {pages} {414001} (\bibinfo {year} {2021})}\BibitemShut {NoStop}%
\bibitem [{\citenamefont {Dias}\ \emph {et~al.}(2018{\natexlab{c}})\citenamefont {Dias}, \citenamefont {Araújo},\ and\ \citenamefont {da~Gama}}]{Dias2018}%
  \BibitemOpen
  \bibfield  {author} {\bibinfo {author} {\bibfnamefont {C.~S.}\ \bibnamefont {Dias}}, \bibinfo {author} {\bibfnamefont {N.~A.~M.}\ \bibnamefont {Araújo}}, \ and\ \bibinfo {author} {\bibfnamefont {M.~M.~T.}\ \bibnamefont {da~Gama}},\ }\href@noop {} {\bibfield  {journal} {\bibinfo  {journal} {J. Phys.: Condens. Matter}\ }\textbf {\bibinfo {volume} {30}},\ \bibinfo {pages} {014001} (\bibinfo {year} {2018}{\natexlab{c}})}\BibitemShut {NoStop}%
\bibitem [{\citenamefont {Cui}\ \emph {et~al.}(2023)\citenamefont {Cui}, \citenamefont {Wang}, \citenamefont {Liang},\ and\ \citenamefont {Qiu}}]{Cui2023}%
  \BibitemOpen
  \bibfield  {author} {\bibinfo {author} {\bibfnamefont {Y.}~\bibnamefont {Cui}}, \bibinfo {author} {\bibfnamefont {J.}~\bibnamefont {Wang}}, \bibinfo {author} {\bibfnamefont {J.}~\bibnamefont {Liang}}, \ and\ \bibinfo {author} {\bibfnamefont {H.}~\bibnamefont {Qiu}},\ }\href@noop {} {\bibfield  {journal} {\bibinfo  {journal} {Small}\ }\textbf {\bibinfo {volume} {19}},\ \bibinfo {pages} {2207609} (\bibinfo {year} {2023})}\BibitemShut {NoStop}%
\bibitem [{\citenamefont {Lee}\ \emph {et~al.}(2011)\citenamefont {Lee}, \citenamefont {Yoon},\ and\ \citenamefont {Lahann}}]{Lee2011}%
  \BibitemOpen
  \bibfield  {author} {\bibinfo {author} {\bibfnamefont {K.~J.}\ \bibnamefont {Lee}}, \bibinfo {author} {\bibfnamefont {J.}~\bibnamefont {Yoon}}, \ and\ \bibinfo {author} {\bibfnamefont {J.}~\bibnamefont {Lahann}},\ }\href@noop {} {\bibfield  {journal} {\bibinfo  {journal} {Curr. Op. Coll. Interf. Sci.}\ }\textbf {\bibinfo {volume} {16}},\ \bibinfo {pages} {195} (\bibinfo {year} {2011})}\BibitemShut {NoStop}%
\bibitem [{\citenamefont {Chang}\ \emph {et~al.}(2021)\citenamefont {Chang}, \citenamefont {Ouhajji}, \citenamefont {Townsend}, \citenamefont {Lacina}, \citenamefont {van Ravensteijn},\ and\ \citenamefont {Kegel}}]{Chang2021}%
  \BibitemOpen
  \bibfield  {author} {\bibinfo {author} {\bibfnamefont {F.}~\bibnamefont {Chang}}, \bibinfo {author} {\bibfnamefont {S.}~\bibnamefont {Ouhajji}}, \bibinfo {author} {\bibfnamefont {A.}~\bibnamefont {Townsend}}, \bibinfo {author} {\bibfnamefont {K.~S.}\ \bibnamefont {Lacina}}, \bibinfo {author} {\bibfnamefont {B.~G.~P.}\ \bibnamefont {van Ravensteijn}}, \ and\ \bibinfo {author} {\bibfnamefont {W.~K.}\ \bibnamefont {Kegel}},\ }\href@noop {} {\bibfield  {journal} {\bibinfo  {journal} {J. Colloid Interface Sci.}\ }\textbf {\bibinfo {volume} {582}},\ \bibinfo {pages} {333} (\bibinfo {year} {2021})}\BibitemShut {NoStop}%
\bibitem [{\citenamefont {Russo}\ and\ \citenamefont {Lattuada}(2024)}]{Russo2024}%
  \BibitemOpen
  \bibfield  {author} {\bibinfo {author} {\bibfnamefont {G.}~\bibnamefont {Russo}}\ and\ \bibinfo {author} {\bibfnamefont {M.}~\bibnamefont {Lattuada}},\ }\href@noop {} {\bibfield  {journal} {\bibinfo  {journal} {Colloids Surf. A}\ }\textbf {\bibinfo {volume} {685}},\ \bibinfo {pages} {133293} (\bibinfo {year} {2024})}\BibitemShut {NoStop}%
\bibitem [{\citenamefont {Tsurusawa}\ \emph {et~al.}(2020)\citenamefont {Tsurusawa}, \citenamefont {Arai},\ and\ \citenamefont {Tanaka}}]{Tsurusawa2020}%
  \BibitemOpen
  \bibfield  {author} {\bibinfo {author} {\bibfnamefont {H.}~\bibnamefont {Tsurusawa}}, \bibinfo {author} {\bibfnamefont {S.}~\bibnamefont {Arai}}, \ and\ \bibinfo {author} {\bibfnamefont {H.}~\bibnamefont {Tanaka}},\ }\href@noop {} {\bibfield  {journal} {\bibinfo  {journal} {Sci. Adv.}\ }\textbf {\bibinfo {volume} {6}},\ \bibinfo {pages} {eabb8107} (\bibinfo {year} {2020})}\BibitemShut {NoStop}%
\bibitem [{\citenamefont {Plimpton}(1995)}]{Plimpton1995}%
  \BibitemOpen
  \bibfield  {author} {\bibinfo {author} {\bibfnamefont {S.}~\bibnamefont {Plimpton}},\ }\href@noop {} {\bibfield  {journal} {\bibinfo  {journal} {J. Comp. Phys.}\ }\textbf {\bibinfo {volume} {117}},\ \bibinfo {pages} {1} (\bibinfo {year} {1995})}\BibitemShut {NoStop}%
\bibitem [{\citenamefont {Mazza}\ \emph {et~al.}(2007)\citenamefont {Mazza}, \citenamefont {Giovambattista}, \citenamefont {Stanley},\ and\ \citenamefont {Starr}}]{Mazza2007}%
  \BibitemOpen
  \bibfield  {author} {\bibinfo {author} {\bibfnamefont {M.~G.}\ \bibnamefont {Mazza}}, \bibinfo {author} {\bibfnamefont {N.}~\bibnamefont {Giovambattista}}, \bibinfo {author} {\bibfnamefont {H.~E.}\ \bibnamefont {Stanley}}, \ and\ \bibinfo {author} {\bibfnamefont {F.~W.}\ \bibnamefont {Starr}},\ }\href@noop {} {\bibfield  {journal} {\bibinfo  {journal} {Phys. Rev. E}\ }\textbf {\bibinfo {volume} {76}},\ \bibinfo {pages} {031203} (\bibinfo {year} {2007})}\BibitemShut {NoStop}%
\bibitem [{\citenamefont {Colombo}\ and\ \citenamefont {Gado}(2014{\natexlab{b}})}]{Colombo2014}%
  \BibitemOpen
  \bibfield  {author} {\bibinfo {author} {\bibfnamefont {J.}~\bibnamefont {Colombo}}\ and\ \bibinfo {author} {\bibfnamefont {E.~D.}\ \bibnamefont {Gado}},\ }\href@noop {} {\bibfield  {journal} {\bibinfo  {journal} {J. Rheol.}\ }\textbf {\bibinfo {volume} {58}},\ \bibinfo {pages} {1089} (\bibinfo {year} {2014}{\natexlab{b}})}\BibitemShut {NoStop}%
\bibitem [{\citenamefont {Rubinstein}\ and\ \citenamefont {Colby}(2003)}]{Rubinstein2003}%
  \BibitemOpen
  \bibfield  {author} {\bibinfo {author} {\bibfnamefont {M.}~\bibnamefont {Rubinstein}}\ and\ \bibinfo {author} {\bibfnamefont {R.~H.}\ \bibnamefont {Colby}},\ }\href@noop {} {\emph {\bibinfo {title} {Polymer Physics}}}\ (\bibinfo  {publisher} {Oxford University PressOxford},\ \bibinfo {year} {2003})\BibitemShut {NoStop}%
\bibitem [{\citenamefont {Bianchi}\ \emph {et~al.}(2006)\citenamefont {Bianchi}, \citenamefont {Largo}, \citenamefont {Tartaglia}, \citenamefont {Zaccarelli},\ and\ \citenamefont {Sciortino}}]{Bianchi2006}%
  \BibitemOpen
  \bibfield  {author} {\bibinfo {author} {\bibfnamefont {E.}~\bibnamefont {Bianchi}}, \bibinfo {author} {\bibfnamefont {J.}~\bibnamefont {Largo}}, \bibinfo {author} {\bibfnamefont {P.}~\bibnamefont {Tartaglia}}, \bibinfo {author} {\bibfnamefont {E.}~\bibnamefont {Zaccarelli}}, \ and\ \bibinfo {author} {\bibfnamefont {F.}~\bibnamefont {Sciortino}},\ }\href@noop {} {\bibfield  {journal} {\bibinfo  {journal} {Phys. Rev. Lett.}\ }\textbf {\bibinfo {volume} {97}},\ \bibinfo {pages} {168301} (\bibinfo {year} {2006})}\BibitemShut {NoStop}%
\bibitem [{\citenamefont {Teixeira}\ and\ \citenamefont {Tavares}(2017)}]{Teixeira2017a}%
  \BibitemOpen
  \bibfield  {author} {\bibinfo {author} {\bibfnamefont {P.~I.~C.}\ \bibnamefont {Teixeira}}\ and\ \bibinfo {author} {\bibfnamefont {J.~M.}\ \bibnamefont {Tavares}},\ }\href@noop {} {\bibfield  {journal} {\bibinfo  {journal} {Curr. Op. Coll. Interf. Sci.}\ }\textbf {\bibinfo {volume} {30}},\ \bibinfo {pages} {16} (\bibinfo {year} {2017})}\BibitemShut {NoStop}%
\bibitem [{\citenamefont {de~las Heras}\ \emph {et~al.}(2011{\natexlab{b}})\citenamefont {de~las Heras}, \citenamefont {Tavares},\ and\ \citenamefont {da~Gama}}]{DelasHeras2011}%
  \BibitemOpen
  \bibfield  {author} {\bibinfo {author} {\bibfnamefont {D.}~\bibnamefont {de~las Heras}}, \bibinfo {author} {\bibfnamefont {J.~M.}\ \bibnamefont {Tavares}}, \ and\ \bibinfo {author} {\bibfnamefont {M.~M.~T.}\ \bibnamefont {da~Gama}},\ }\href@noop {} {\bibfield  {journal} {\bibinfo  {journal} {Soft Matt.}\ }\textbf {\bibinfo {volume} {7}},\ \bibinfo {pages} {5615} (\bibinfo {year} {2011}{\natexlab{b}})}\BibitemShut {NoStop}%
\bibitem [{\citenamefont {Gouveia}\ \emph {et~al.}(2022)\citenamefont {Gouveia}, \citenamefont {Dias},\ and\ \citenamefont {Tavares}}]{Gouveia2022}%
  \BibitemOpen
  \bibfield  {author} {\bibinfo {author} {\bibfnamefont {M.}~\bibnamefont {Gouveia}}, \bibinfo {author} {\bibfnamefont {C.~S.}\ \bibnamefont {Dias}}, \ and\ \bibinfo {author} {\bibfnamefont {J.~M.}\ \bibnamefont {Tavares}},\ }\href@noop {} {\bibfield  {journal} {\bibinfo  {journal} {J. Chem. Phys.}\ }\textbf {\bibinfo {volume} {157}},\ \bibinfo {pages} {164903} (\bibinfo {year} {2022})}\BibitemShut {NoStop}%
\bibitem [{\citenamefont {Russo}\ \emph {et~al.}(2011)\citenamefont {Russo}, \citenamefont {Tavares}, \citenamefont {Teixeira}, \citenamefont {da~Gama},\ and\ \citenamefont {Sciortino}}]{Russo2011a}%
  \BibitemOpen
  \bibfield  {author} {\bibinfo {author} {\bibfnamefont {J.}~\bibnamefont {Russo}}, \bibinfo {author} {\bibfnamefont {J.~M.}\ \bibnamefont {Tavares}}, \bibinfo {author} {\bibfnamefont {P.~I.~C.}\ \bibnamefont {Teixeira}}, \bibinfo {author} {\bibfnamefont {M.~M.~T.}\ \bibnamefont {da~Gama}}, \ and\ \bibinfo {author} {\bibfnamefont {F.}~\bibnamefont {Sciortino}},\ }\href@noop {} {\bibfield  {journal} {\bibinfo  {journal} {Phys. Rev. Lett.}\ }\textbf {\bibinfo {volume} {106}},\ \bibinfo {pages} {085703} (\bibinfo {year} {2011})}\BibitemShut {NoStop}%
\bibitem [{\citenamefont {de~Las~Heras}\ \emph {et~al.}(2012)\citenamefont {de~Las~Heras}, \citenamefont {Tavares},\ and\ \citenamefont {da~Gama}}]{DelasHeras2012}%
  \BibitemOpen
  \bibfield  {author} {\bibinfo {author} {\bibfnamefont {D.}~\bibnamefont {de~Las~Heras}}, \bibinfo {author} {\bibfnamefont {J.~M.}\ \bibnamefont {Tavares}}, \ and\ \bibinfo {author} {\bibfnamefont {M.~M.~T.}\ \bibnamefont {da~Gama}},\ }\href@noop {} {\bibfield  {journal} {\bibinfo  {journal} {Soft Matt.}\ }\textbf {\bibinfo {volume} {8}},\ \bibinfo {pages} {1785} (\bibinfo {year} {2012})}\BibitemShut {NoStop}%
\bibitem [{\citenamefont {Lei}\ \emph {et~al.}(2022)\citenamefont {Lei}, \citenamefont {Bai}, \citenamefont {Qin}, \citenamefont {Liu}, \citenamefont {Huang},\ and\ \citenamefont {Lv}}]{Lei2022}%
  \BibitemOpen
  \bibfield  {author} {\bibinfo {author} {\bibfnamefont {L.}~\bibnamefont {Lei}}, \bibinfo {author} {\bibfnamefont {Y.}~\bibnamefont {Bai}}, \bibinfo {author} {\bibfnamefont {X.}~\bibnamefont {Qin}}, \bibinfo {author} {\bibfnamefont {J.}~\bibnamefont {Liu}}, \bibinfo {author} {\bibfnamefont {W.}~\bibnamefont {Huang}}, \ and\ \bibinfo {author} {\bibfnamefont {Q.}~\bibnamefont {Lv}},\ }\href@noop {} {\bibfield  {journal} {\bibinfo  {journal} {Gels}\ }\textbf {\bibinfo {volume} {8}},\ \bibinfo {pages} {301} (\bibinfo {year} {2022})}\BibitemShut {NoStop}%
\bibitem [{\citenamefont {Custódio}\ \emph {et~al.}(2015)\citenamefont {Custódio}, \citenamefont {Cerqueira}, \citenamefont {Marques}, \citenamefont {Reis},\ and\ \citenamefont {Mano}}]{Custodio2015}%
  \BibitemOpen
  \bibfield  {author} {\bibinfo {author} {\bibfnamefont {C.~A.}\ \bibnamefont {Custódio}}, \bibinfo {author} {\bibfnamefont {M.~T.}\ \bibnamefont {Cerqueira}}, \bibinfo {author} {\bibfnamefont {A.~P.}\ \bibnamefont {Marques}}, \bibinfo {author} {\bibfnamefont {R.~L.}\ \bibnamefont {Reis}}, \ and\ \bibinfo {author} {\bibfnamefont {J.~F.}\ \bibnamefont {Mano}},\ }\href@noop {} {\bibfield  {journal} {\bibinfo  {journal} {Biomaterials}\ }\textbf {\bibinfo {volume} {43}},\ \bibinfo {pages} {23} (\bibinfo {year} {2015})}\BibitemShut {NoStop}%
\bibitem [{\citenamefont {Custódio}\ \emph {et~al.}(2014)\citenamefont {Custódio}, \citenamefont {Santo}, \citenamefont {Oliveira}, \citenamefont {Gomes}, \citenamefont {Reis},\ and\ \citenamefont {Mano}}]{Custodio2014}%
  \BibitemOpen
  \bibfield  {author} {\bibinfo {author} {\bibfnamefont {C.~A.}\ \bibnamefont {Custódio}}, \bibinfo {author} {\bibfnamefont {V.~E.}\ \bibnamefont {Santo}}, \bibinfo {author} {\bibfnamefont {M.~B.}\ \bibnamefont {Oliveira}}, \bibinfo {author} {\bibfnamefont {M.~E.}\ \bibnamefont {Gomes}}, \bibinfo {author} {\bibfnamefont {R.~L.}\ \bibnamefont {Reis}}, \ and\ \bibinfo {author} {\bibfnamefont {J.~F.}\ \bibnamefont {Mano}},\ }\href@noop {} {\bibfield  {journal} {\bibinfo  {journal} {Adv. Funct. Mater.}\ }\textbf {\bibinfo {volume} {24}},\ \bibinfo {pages} {1391} (\bibinfo {year} {2014})}\BibitemShut {NoStop}%
\bibitem [{\citenamefont {Belly}\ \emph {et~al.}(2022)\citenamefont {Belly}, \citenamefont {Paluch},\ and\ \citenamefont {Chalut}}]{DeBelly2022}%
  \BibitemOpen
  \bibfield  {author} {\bibinfo {author} {\bibfnamefont {H.~D.}\ \bibnamefont {Belly}}, \bibinfo {author} {\bibfnamefont {E.~K.}\ \bibnamefont {Paluch}}, \ and\ \bibinfo {author} {\bibfnamefont {K.~J.}\ \bibnamefont {Chalut}},\ }\href@noop {} {\bibfield  {journal} {\bibinfo  {journal} {Nat. Rev. Mol. Cell. Biol.}\ }\textbf {\bibinfo {volume} {23}},\ \bibinfo {pages} {465} (\bibinfo {year} {2022})}\BibitemShut {NoStop}%
\bibitem [{\citenamefont {Sanz-Herrera}\ \emph {et~al.}(2009)\citenamefont {Sanz-Herrera}, \citenamefont {Moreo}, \citenamefont {García-Aznar},\ and\ \citenamefont {Doblaré}}]{Sanz-Herrera2009}%
  \BibitemOpen
  \bibfield  {author} {\bibinfo {author} {\bibfnamefont {J.~A.}\ \bibnamefont {Sanz-Herrera}}, \bibinfo {author} {\bibfnamefont {P.}~\bibnamefont {Moreo}}, \bibinfo {author} {\bibfnamefont {J.~M.}\ \bibnamefont {García-Aznar}}, \ and\ \bibinfo {author} {\bibfnamefont {M.}~\bibnamefont {Doblaré}},\ }\href@noop {} {\bibfield  {journal} {\bibinfo  {journal} {Biomaterials}\ }\textbf {\bibinfo {volume} {30}},\ \bibinfo {pages} {6674} (\bibinfo {year} {2009})}\BibitemShut {NoStop}%
\bibitem [{\citenamefont {Vassaux}\ and\ \citenamefont {Milan}(2017)}]{Vassaux2017}%
  \BibitemOpen
  \bibfield  {author} {\bibinfo {author} {\bibfnamefont {M.}~\bibnamefont {Vassaux}}\ and\ \bibinfo {author} {\bibfnamefont {J.~L.}\ \bibnamefont {Milan}},\ }\href@noop {} {\bibfield  {journal} {\bibinfo  {journal} {Biomech. Model Mechanobiol.}\ }\textbf {\bibinfo {volume} {16}},\ \bibinfo {pages} {1295} (\bibinfo {year} {2017})}\BibitemShut {NoStop}%
\bibitem [{\citenamefont {Gudipaty}\ \emph {et~al.}(2017)\citenamefont {Gudipaty}, \citenamefont {Lindblom}, \citenamefont {Loftus}, \citenamefont {Redd}, \citenamefont {Edes}, \citenamefont {Davey}, \citenamefont {Krishnegowda},\ and\ \citenamefont {Rosenblatt}}]{Gudipaty2017}%
  \BibitemOpen
  \bibfield  {author} {\bibinfo {author} {\bibfnamefont {S.~A.}\ \bibnamefont {Gudipaty}}, \bibinfo {author} {\bibfnamefont {J.}~\bibnamefont {Lindblom}}, \bibinfo {author} {\bibfnamefont {P.~D.}\ \bibnamefont {Loftus}}, \bibinfo {author} {\bibfnamefont {M.~J.}\ \bibnamefont {Redd}}, \bibinfo {author} {\bibfnamefont {K.}~\bibnamefont {Edes}}, \bibinfo {author} {\bibfnamefont {C.~F.}\ \bibnamefont {Davey}}, \bibinfo {author} {\bibfnamefont {V.}~\bibnamefont {Krishnegowda}}, \ and\ \bibinfo {author} {\bibfnamefont {J.}~\bibnamefont {Rosenblatt}},\ }\href@noop {} {\bibfield  {journal} {\bibinfo  {journal} {Nature}\ }\textbf {\bibinfo {volume} {543}},\ \bibinfo {pages} {118} (\bibinfo {year} {2017})}\BibitemShut {NoStop}%
\bibitem [{\citenamefont {Luo}\ \emph {et~al.}(2013)\citenamefont {Luo}, \citenamefont {Mohan}, \citenamefont {Iglesias},\ and\ \citenamefont {Robinson}}]{Luo2013}%
  \BibitemOpen
  \bibfield  {author} {\bibinfo {author} {\bibfnamefont {T.}~\bibnamefont {Luo}}, \bibinfo {author} {\bibfnamefont {K.}~\bibnamefont {Mohan}}, \bibinfo {author} {\bibfnamefont {P.~A.}\ \bibnamefont {Iglesias}}, \ and\ \bibinfo {author} {\bibfnamefont {D.~N.}\ \bibnamefont {Robinson}},\ }\href@noop {} {\bibfield  {journal} {\bibinfo  {journal} {Nat. Mater.}\ }\textbf {\bibinfo {volume} {12}},\ \bibinfo {pages} {1064} (\bibinfo {year} {2013})}\BibitemShut {NoStop}%
\bibitem [{\citenamefont {Marinari}\ \emph {et~al.}(2012)\citenamefont {Marinari}, \citenamefont {Mehonic}, \citenamefont {Curran}, \citenamefont {Gale}, \citenamefont {Duke},\ and\ \citenamefont {Baum}}]{Marinari2012}%
  \BibitemOpen
  \bibfield  {author} {\bibinfo {author} {\bibfnamefont {E.}~\bibnamefont {Marinari}}, \bibinfo {author} {\bibfnamefont {A.}~\bibnamefont {Mehonic}}, \bibinfo {author} {\bibfnamefont {S.}~\bibnamefont {Curran}}, \bibinfo {author} {\bibfnamefont {J.}~\bibnamefont {Gale}}, \bibinfo {author} {\bibfnamefont {T.}~\bibnamefont {Duke}}, \ and\ \bibinfo {author} {\bibfnamefont {B.}~\bibnamefont {Baum}},\ }\href@noop {} {\bibfield  {journal} {\bibinfo  {journal} {Nature}\ }\textbf {\bibinfo {volume} {484}},\ \bibinfo {pages} {542} (\bibinfo {year} {2012})}\BibitemShut {NoStop}%
\bibitem [{\citenamefont {Mitura}\ \emph {et~al.}(2020)\citenamefont {Mitura}, \citenamefont {Sionkowska},\ and\ \citenamefont {Jaiswal}}]{Mitura2020}%
  \BibitemOpen
  \bibfield  {author} {\bibinfo {author} {\bibfnamefont {S.}~\bibnamefont {Mitura}}, \bibinfo {author} {\bibfnamefont {A.}~\bibnamefont {Sionkowska}}, \ and\ \bibinfo {author} {\bibfnamefont {A.}~\bibnamefont {Jaiswal}},\ }\href@noop {} {\bibfield  {journal} {\bibinfo  {journal} {J. Mater. Sci.: Mater. Med.}\ }\textbf {\bibinfo {volume} {31}},\ \bibinfo {pages} {50} (\bibinfo {year} {2020})}\BibitemShut {NoStop}%
\bibitem [{\citenamefont {Ioannidou}\ \emph {et~al.}(2016)\citenamefont {Ioannidou}, \citenamefont {Kanduč}, \citenamefont {Li}, \citenamefont {Frenkel}, \citenamefont {Dobnikar},\ and\ \citenamefont {Gado}}]{Ioannidou2016}%
  \BibitemOpen
  \bibfield  {author} {\bibinfo {author} {\bibfnamefont {K.}~\bibnamefont {Ioannidou}}, \bibinfo {author} {\bibfnamefont {M.}~\bibnamefont {Kanduč}}, \bibinfo {author} {\bibfnamefont {L.}~\bibnamefont {Li}}, \bibinfo {author} {\bibfnamefont {D.}~\bibnamefont {Frenkel}}, \bibinfo {author} {\bibfnamefont {J.}~\bibnamefont {Dobnikar}}, \ and\ \bibinfo {author} {\bibfnamefont {E.~D.}\ \bibnamefont {Gado}},\ }\href@noop {} {\bibfield  {journal} {\bibinfo  {journal} {Nat. Commun.}\ }\textbf {\bibinfo {volume} {7}},\ \bibinfo {pages} {12106} (\bibinfo {year} {2016})}\BibitemShut {NoStop}%
\bibitem [{\citenamefont {Banerjee}\ and\ \citenamefont {Bhattacharya}(2012)}]{Banerjee2012}%
  \BibitemOpen
  \bibfield  {author} {\bibinfo {author} {\bibfnamefont {S.}~\bibnamefont {Banerjee}}\ and\ \bibinfo {author} {\bibfnamefont {S.}~\bibnamefont {Bhattacharya}},\ }\href@noop {} {\bibfield  {journal} {\bibinfo  {journal} {Crit. Rev. Food Sci. Nut.}\ }\textbf {\bibinfo {volume} {52}},\ \bibinfo {pages} {334} (\bibinfo {year} {2012})}\BibitemShut {NoStop}%
\bibitem [{\citenamefont {Coropceanu}\ \emph {et~al.}(2022)\citenamefont {Coropceanu}, \citenamefont {Janke}, \citenamefont {Portner}, \citenamefont {Haubold}, \citenamefont {Nguyen}, \citenamefont {Das}, \citenamefont {Tanner}, \citenamefont {Utterback}, \citenamefont {Teitelbaum}, \citenamefont {Hudson}, \citenamefont {Sarma}, \citenamefont {Hinkle}, \citenamefont {Tassone}, \citenamefont {Eychmüller}, \citenamefont {Limmer}, \citenamefont {de~la Cruz}, \citenamefont {Ginsberg},\ and\ \citenamefont {Talapin}}]{Coropceanu2022}%
  \BibitemOpen
  \bibfield  {author} {\bibinfo {author} {\bibfnamefont {I.}~\bibnamefont {Coropceanu}}, \bibinfo {author} {\bibfnamefont {E.~M.}\ \bibnamefont {Janke}}, \bibinfo {author} {\bibfnamefont {J.}~\bibnamefont {Portner}}, \bibinfo {author} {\bibfnamefont {D.}~\bibnamefont {Haubold}}, \bibinfo {author} {\bibfnamefont {T.~D.}\ \bibnamefont {Nguyen}}, \bibinfo {author} {\bibfnamefont {A.}~\bibnamefont {Das}}, \bibinfo {author} {\bibfnamefont {C.~P.~N.}\ \bibnamefont {Tanner}}, \bibinfo {author} {\bibfnamefont {J.~K.}\ \bibnamefont {Utterback}}, \bibinfo {author} {\bibfnamefont {S.~W.}\ \bibnamefont {Teitelbaum}}, \bibinfo {author} {\bibfnamefont {M.~H.}\ \bibnamefont {Hudson}}, \bibinfo {author} {\bibfnamefont {N.~A.}\ \bibnamefont {Sarma}}, \bibinfo {author} {\bibfnamefont {A.~M.}\ \bibnamefont {Hinkle}}, \bibinfo {author} {\bibfnamefont {C.~J.}\ \bibnamefont {Tassone}}, \bibinfo {author} {\bibfnamefont {A.}~\bibnamefont {Eychmüller}}, \bibinfo {author} {\bibfnamefont {D.~T.}\ \bibnamefont {Limmer}}, \bibinfo
  {author} {\bibfnamefont {M.~O.}\ \bibnamefont {de~la Cruz}}, \bibinfo {author} {\bibfnamefont {N.~S.}\ \bibnamefont {Ginsberg}}, \ and\ \bibinfo {author} {\bibfnamefont {D.~V.}\ \bibnamefont {Talapin}},\ }\href@noop {} {\bibfield  {journal} {\bibinfo  {journal} {Science}\ }\textbf {\bibinfo {volume} {375}},\ \bibinfo {pages} {1422} (\bibinfo {year} {2022})}\BibitemShut {NoStop}%
\end{thebibliography}%

\end{document}